\documentclass[a4paper,fleqn,usenatbib]{mnras}
\usepackage{color}
\usepackage{graphicx}
\usepackage{courier}
\usepackage{subcaption}
\usepackage{caption}
\usepackage{amsmath}
\usepackage{longtable}
\usepackage{sidecap}
\usepackage{multicol}
\usepackage{floatrow}
\usepackage{newtxtext,newtxmath}

\usepackage[T1]{fontenc}
\usepackage{ae,aecompl}

\title[Core Properties of the BAT AGN]{BAT AGN Spectroscopic Survey - XV: The High Frequency Radio Cores of Ultra-hard X-ray Selected AGN}

\author[K. L. Smith et al.]{
Krista Lynne Smith,$^{1,2}$\thanks{E-mail: klynne@stanford.edu} 
Richard F. Mushotzky,$^{3}$ 
Michael Koss,$^{4}$ 
Benny Trakhtenbrot,$^{5}$ 
\newauthor
Claudio Ricci,$^{6,7}$ 
O. Ivy Wong,$^{8}$ 
Franz E. Bauer,$^{6,9,10}$ 
Federica Ricci,$^{6}$ 
Stuart Vogel,$^{3}$
\newauthor
Daniel Stern,$^{11}$ 
Meredith C. Powell,$^{12}$ 
C. Meg Urry,$^{12}$ 
Fiona Harrison,$^{13}$ 
\newauthor
Julian Mejia-Restrepo,$^{14}$ 
Kyuseok Oh,$^{15,16,17}$ 
Junhyun Baek,$^{18}$ 
and Aeree Chung$^{18}$
\\
$^{1}$KIPAC at SLAC, Stanford University, Menlo Park CA 94025, klynne@stanford.edu\\
$^{2}$Einstein Fellow\\
$^{3}$Department of Astronomy, University of Maryland, College Park, MD 20742\\
$^{4}$Eureka Scientific Inc., 2452 Delmer St. Suite 100, Oakland, CA 94602\\
$^{5}$School of Physics and Astronomy, Tel Aviv University, Tel Aviv 69978, Israel\\
$^{6}$Instituto de Astrof{\'{\i}}sica and Centro de Astroingenier{\'{\i}}a, Facultad de F{\'{i}}sica, Pontificia Universidad Cat{\'{o}}lica de Chile, Casilla 306, Santiago 22, Chile\\ 
$^{7}$Kavli Institute for Astronomy and Astrophysics, Peking University, Beijing 100871, China\\
$^{8}$ICRAR, The University of Western Australia, Crawley, Perth, Western Australia, 6009\\
$^{9}$Millennium Institute of Astrophysics (MAS), Nuncio Monse{\~{n}} or S{\'{o}}tero Sanz 100, Providencia, Santiago, Chile\\
$^{10}$Space Science Institute, 4750 Walnut Street, Suite 205, Boulder, Colorado 80301\\
$^{11}$Jet Propulsion Laboratory, California Institute of Technology, 4800 Oak Grove Drive, MS 169-224, Pasadena, CA 91109, USA\\
$^{12}$Yale Center for Astronomy and Astrophysics, and Physics Department, Yale University, P.O. Box 2018120, New Haven, CT 06520-8120\\
$^{13}$Cahill Center for Astronomy and Astrophysics, California Institute of Technology, Pasadena, CA 91125, USA\\
$^{14}$European Southern Observatory, Alonso de Cordova 3107, Casilla 19001, Victacura, Santiago, Chile\\
$^{15}$Department of Astronomy, Kyoto University, Kitashirakawa-Oiwake-cho, Sakyo-ku, Kyoto 606-8502, Japan\\
$^{16}$JSPS Fellow\\
$^{17}$Korea Astronomy \& Space Science institute, 776, Daedeokdae-ro, Yuseong-gu, Daejeon 34055, Republic of Korea\\
$^{18}$Department of Astronomy, Yonsei University, 50 Yonsei-ro, Seodaemun-gu, Seoul 03722, Republic of Korea
}

\date{Accepted XXX. Received YYY; in original form ZZZ}

\pubyear{2019}
\begin{document}
\label{firstpage}
\pagerange{\pageref{firstpage}--\pageref{lastpage}}
\maketitle

\begin{abstract}

We have conducted 22 GHz radio imaging at 1\arcsec~resolution of 100 low-redshift AGN selected at 14-195 keV by the Swift-BAT. We find a radio core detection fraction of 96\%, much higher than lower-frequency radio surveys. Of the 96 radio-detected AGN, 55 have compact morphologies, 30 have morphologies consistent with nuclear star formation, and 11 have sub-kpc to kpc-scale jets. We find that the total radio power does not distinguish between nuclear star formation and jets as the origin of the radio emission. For 87 objects, we use optical spectroscopy to test whether AGN physical parameters are distinct between radio morphological types. We find that X-ray luminosities tend to be higher if the 22 GHz morphology is jet-like, but find no significant difference in other physical parameters. We find that the relationship between the X-ray and core radio luminosities is consistent with the $L_R/L_X \sim 10^{-5}$ of coronally active stars. We further find that the canonical fundamental planes of black hole activity systematically over-predict our radio luminosities, particularly for objects with star formation morphologies.

\end{abstract}


\section{Introduction}
\label{sec:intro}

Active galactic nuclei (AGN) are powered by accretion onto supermassive black holes and emit
strongly across the entire electromagnetic spectrum. Some AGN exhibit powerful radio jets that extend well outside the host galaxy, with dramatic effects on the host itself and the surrounding medium. These objects, however, are the exception. The vast majority of AGN are radio-quiet. Such AGN may exhibit either an unresolved radio core, or an unresolved core and local, extended emission that may be related to outflows or star formation \citep[e.g.,][]{Ulvestad1984,Edelson1987,Giuricin1990,Nagar1999,Orienti2010}; see also the recent review by \citet{Panessa2019}. Additionally, many radio-quiet AGN have stubbornly refused detection at all, despite surveys at a variety of observing frequencies and resolutions \citep[e.g.,][]{Roy1998,Maini2016,Herrera-Ruiz2016}

A key question is whether the radio emission mechanism in the core is the same in objects with and without powerful radio jets. One prevalent idea is that the same mechanism responsible for the powerful jets operates in a scale-invariant way down to the faintest radio luminosities, and that the radio emission is simply due to smaller, unresolved jets \citep[e.g.,][]{Miller1993,Heinz2003}. This conclusion is supported by the discovery of the ``fundamental plane of black hole activity," a remarkably tight relationship between the X-ray luminosity, radio luminosity, and black hole mass that appears to apply to both stellar mass and supermassive black holes \citep{Merloni2003,Falcke2004,Kording2006}. 

However, very high-resolution radio imaging of radio-quiet quasars do not always find jets; although VLBI imaging campaigns revealed sub-parsec scale jets in some radio-quiet Seyferts, many remain unresolved \citep{Ulvestad2001,Ulvestad2003,Ulvestad2005}. These same studies concluded that thermal emission or low-efficiency accretion scenarios were ruled out by the very high implied brightness temperatures, and theorized that perhaps radio-quiet quasars simply produce weaker jets that are disrupted by passage through the host galaxy, and are too faint to be detected in all but the deepest observations.

An alternative proposal is that instead of unresolved jets, the core radio emission in radio-quiet objects is dominated by a coronal component in which both radio and X-ray emission is generated in a region of hot plasma associated with the accretion flow \citep{Laor2008,Raginski2016}. 

For the past two years, we have conducted a 22~GHz radio imaging survey at 1\arcsec~resolution of a low-redshift, largely radio-quiet subset of the ultra-hard X-ray selected \emph{Swift}-BAT AGN sample \citep{Baumgartner2013}. In this work, we use this unique sample to place new constraints on the origin of radio emission in AGN without powerful radio jets by providing a 22~GHz radio detection fraction and comparing the results to the coronal $L_R/L_X$ relation and existing fundamental plane relations.

 Key to enabling the expansion of this investigation into the physical origin of the radio emission is the BAT AGN Spectroscopic Survey (BASS; \citealt{Koss2017}, \citealt{Ricci2017}), a large, ongoing collaborative multiwavelength effort to obtain spectra and imaging of the BAT AGN. 

In Section~\ref{sec:sample}, we present the sample selection and analysis of the JVLA radio data and the supplementary parameter measurements from the BASS. In Section~\ref{sec:morph}, we describe the radio morphologies in our survey. Section~\ref{sec:paramstats} incorporates the black hole masses, accretion rates, spectral indices, and luminosities in a comparative analysis. In Section~\ref{sec:lrlx}, we discuss the correlation between radio and X-ray luminosities in our sample; in Section~\ref{sec:fundplane} we expand this into a discussion of how our sample compares to the fundamental plane of black hole activity. The results of the preceding sections are discussed with scientific context in Section~\ref{sec:discussion}. Conclusions are presented and summarized in Section~\ref{sec:conclusion}. 

Throughout the paper we assume a cosmology of $H_0 = 69.6$~km s$^{-1}$~Mpc$^{-1}$, $\Omega_M = 0.286$, and $\Omega_\Lambda = 0.714$.

\section{Sample Selection and Data Reduction}
\label{sec:sample}
\subsection{Selection and Properties of the JVLA 22~GHz Sample}
\label{sec:radiosample}

The sample presented in this paper was observed in two main stages motivated by different science goals. The spatially resolved core radio emission, the subject of this paper, was an incidental product of the high-resolution imaging required for those goals. In order to clarify the nature of the sample, we describe here the motivations and target selection of the VLA campaigns that yielded the present sample.

Our targets are from the \emph{Swift}-BAT All Sky Survey, conducted in the ultra-hard $14-195$~keV band \citep{Baumgartner2013}. This band is immune to the majority of biases that affect AGN selection in the optical, infrared, and radio, and is sensitive to even highly obscured AGN with column densities as high as $\sim10^{24}$~cm$^{-2}$; see Figure~2 in \citet{Koss2016} for a comparison to other X-ray surveys. Our original parent sample is the 58-month version of the survey, consisting of all 313 X-ray sources identified with low-redshift ($z<0.05$), non-blazar AGN. The sample is composed mainly of Seyfert galaxies with moderate bolometric luminosities $10^{42} < L_\mathrm{bol} < 10^{46}$~erg s$^{-1}$, but also includes a few much rarer luminous quasars. The star formation properties of this sample were studied in the FIR with \emph{Herschel} by \citet{Mushotzky2014}, \citet{Melendez2014}, \citet{Shimizu2015}, and \citet{Shimizu2016}. 

Our initial goal was to study circumnuclear star formation potentially being impacted by the AGN. For this reason, in the first phase of the radio survey we selected objects that were unresolved or only partially resolved in the \emph{Herschel} images. With the declination cut required for JVLA observations, this resulted in 70 objects. The sample was observed at 22~GHz with 1\arcsec~spatial resolution using the JVLA in C~configuration, as described in \citet{Smith2016}. 

One intriguing result derived in \citet{Smith2016} was the discovery of a preponderance of objects with jet-like 22~GHz morphologies in AGN that lay below the ``main-sequence of star formation"  (i.e., with suppressed IR-measured star formation rates compared to normal star forming galaxies for a given total stellar mass). Indeed, \citet{Shimizu2015} had already found that the \emph{Herschel}-observed parent sample lay systematically below the star forming main sequence, in the so-called ``green valley" between star forming galaxies and quiescent ellipticals. Motivated by the possibility that kiloparsec-scale jets might be responsible for star formation suppression, we selected 36 additional AGN from the original parent sample that were at least 1$\sigma$~ below the star forming main sequence to see if the jet preponderance remained. This analysis is described in a concurrent publication (Smith et al. 2020, in preparation). Note that this sample had no cuts made based on whether the \emph{Herschel} images were unresolved.

Together, the total JVLA 22~GHz sample consists of 100 objects after discarding six for persistent radio frequency interference (RFI) and including four non-detections (Mrk~653, Mrk~595, 2MASX~J0107-1139, and Mrk~352). The total redshift range for the observed sample is $0.003\leq z \leq 0.049$, corresponding to spatial beam extents of $62-965$~parsecs. Radio flux densities, X-ray luminosities, and redshifts are given in Table~\ref{t:tab1}. 

The redshift and \emph{Herschel}-resolution cuts removed all radio-loud quasars (including FR~I and FR~II sources) from our sample. This is unsurprising, as such objects are intrinsically rare and tend to be higher redshift; for example, \citet{Gupta2018} found that 51/509 (10\%) of the overall BAT AGN sample are radio loud. The large majority of our remaining sources are therefore radio-quiet Seyferts. However, the boundary between radio-loud and radio-quiet is not well-defined and is also wavelength dependent. To roughly quantify the radio-quietness of our sample we use the \citet{Kellermann1989} quantity $R_O = S_{\nu,\mathrm{5GHz}} / S_B$. We use archival values of the optical $B$ magnitudes or, in the absence of a $B$ value, a $g$ magnitude and convert to flux. To determine the proper radio value, we interpolate between our 22~GHz measurement and archival 1.4~GHz fluxes from the FIRST survey, where they exist. In the absence of a FIRST detection, we assume a radio spectral index of $\alpha=-0.7$ \citep{Kellermann1968, Amirkhanian1985}, where the flux density $S_\nu \sim \nu^\alpha$. Traditionally, $R_O\sim10$ is considered the boundary between radio-loud and radio-quiet objects. Four objects in our sample are near or above this threshold: Arp 102B ($R_O=44$), NGC 5506 ($R_O=13$), Mrk 477 ($R_O=10$), and NGC 1052 ($R_O=9$). Two of these, Arp~102B and NGC~1052, have small radio jets at low frequencies \citep{Helmboldt2007,Wrobel1984}. They are interesting in this context as comparison objects, and are denoted in the figures. 

The optical magnitude of Type 2 AGN will be dominated by host galaxy starlight; in these objects the nuclear optical luminosity is obscured and unknown. Although the radio-loudness $R$ parameter is still often used as a rough estimator, we supplement it here with the X-ray radio loudness criterion from \citet{Terashima2003}: $R_X = \nu L_{\nu,\mathrm{5GHz}} / L_\mathrm{2-10keV}$. There is no bimodality associated with this criterion, which correlates roughly with $R_O$. The optical diagnostic log~$R_O\sim1$ corresponds to a log~$R_X$ value of $-4.5$, the value that they define as the radio-loud boundary; more than half of our objects are classified as radio-loud by this criterion. This is not unusual; \citet{Ulvestad2005} found that their entire sample of radio-quiet quasars were defined as radio-loud by the $R_X$ criterion and many Seyferts and optically-selected quasars in \citet{Terashima2003} itself are above it. 

\subsection{The BAT AGN Spectroscopic Survey}
\label{sec:bass}

Much of the analysis done in this paper makes use of physical parameter estimates from the BAT AGN Spectroscopic Survey (BASS; \citealt{Koss2017}), a large effort to collect optical spectra for the \emph{Swift}-BAT AGN with the goal of leveraging this unbiased sample for black hole mass, accretion rate, and luminosity estimation. In addition to the optical spectroscopic work, the BASS includes careful multi-facility determination of the intrinsic X-ray spectral energy distribution \citep{Ricci2017}. Of the 100 AGN in our survey, 91 are included in the second data release of the spectroscopic survey, of which 82 have black hole mass and Eddington ratio estimates; the black hole mass calculations are discussed in detail in Section~\ref{sec:mbh_calc}. At the time of submission of this paper only Data Release 1 is publicly available; the Data Release 2 products we use here are internal but will shortly be published. 

The optical spectra were obtained from a large variety of telescopes, and can all be viewed at the BASS website\footnote{\url{www.bass-survey.com}}. Although the previous sample of 70 BAT AGN analyzed by \citet{Smith2016} comprises the majority of the targets here and our previous analysis includes many of the same tests presented in that work, the inclusion of the BASS data represents a significant improvement on the tests requiring black hole mass and accretion rate estimates. In \citet{Smith2016}, mass measurements were available for only 16 objects from the literature. The fundamental plane tests especially are far more robust in this experiment due to the additional masses. BASS measurements of black hole mass and accretion rate are provided in Table~\ref{t:tab1}.

\begin{figure*}
    \centering
    \includegraphics[width=0.9\textwidth]{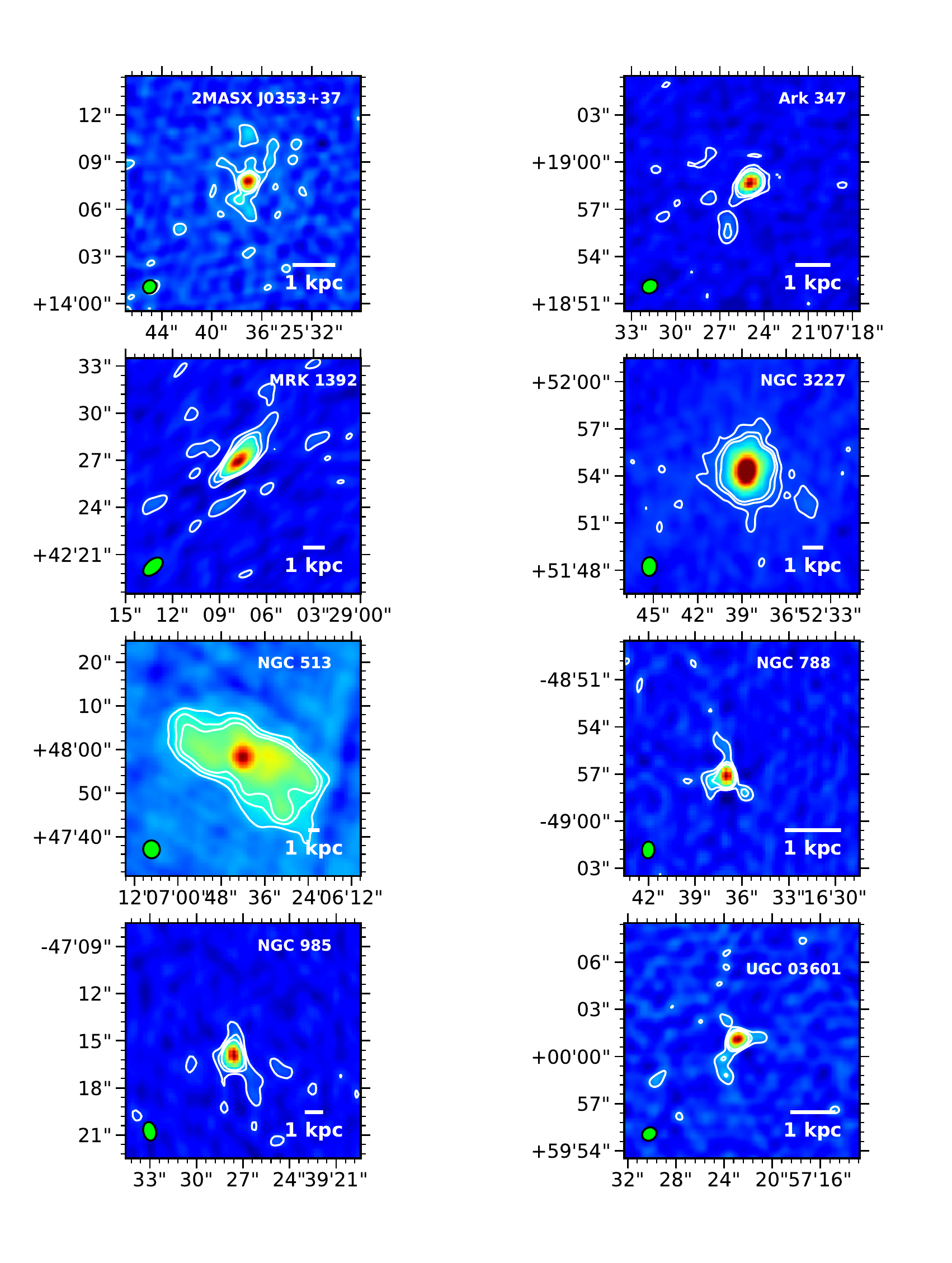}
    \caption{22~GHz images of the BAT AGN with non jet-like morphologies likely due to star formation. Contours occur at 3$\sigma$, 6$\sigma$, and 9$\sigma$ above the background. Each figure includes the beam (green, bottom left) and a scalebar representing 1 kiloparsec. The objects are shown at full 1\arcsec~resolution except for NGC~513, which is shown at 3\arcsec~resolution to best illustrate the extent of the low surface brightness emission.}
    \label{fig:sfmorphs}
\end{figure*}


\begin{figure*}
    \centering
    \includegraphics[width=0.9\textwidth]{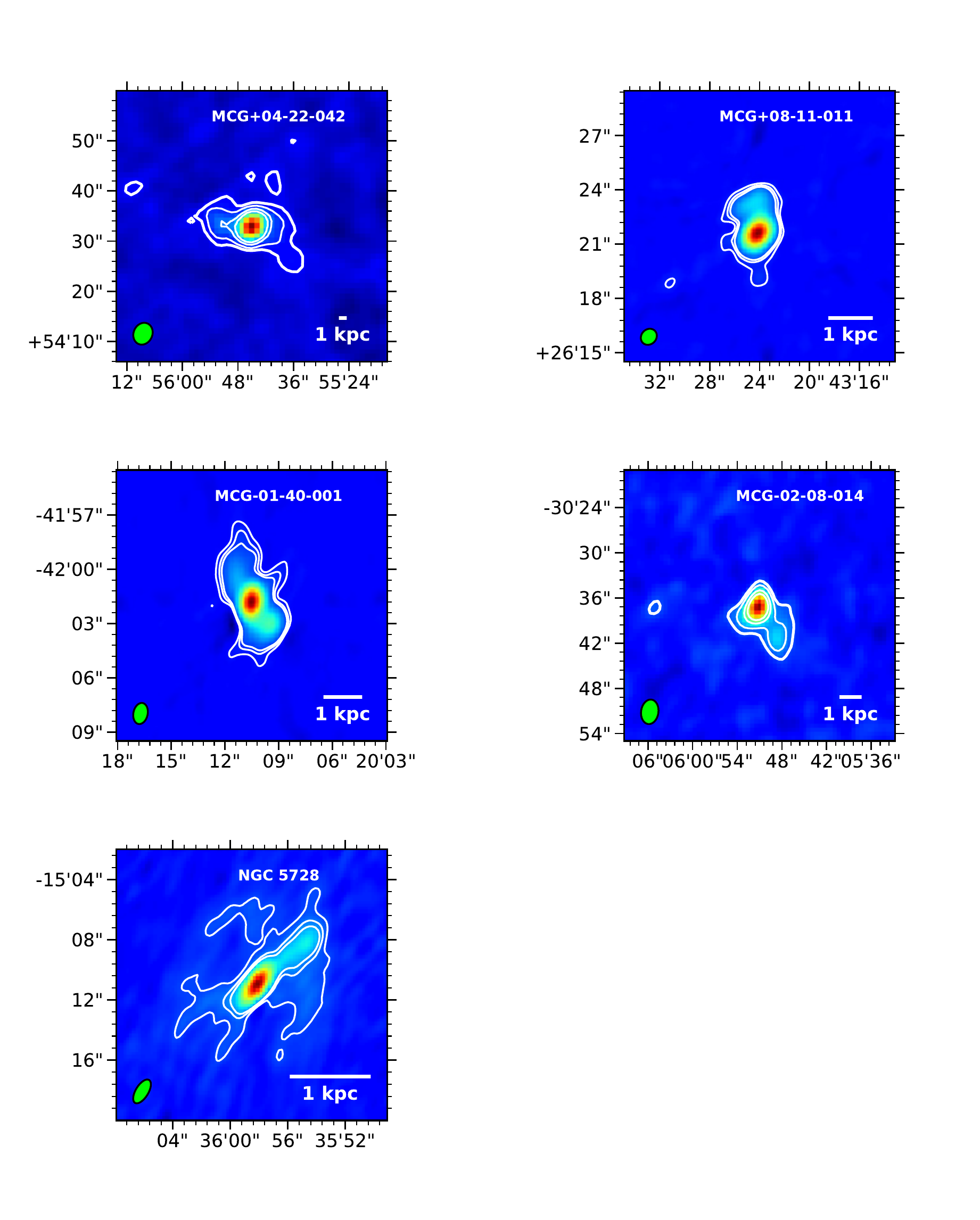}
    \caption{22~GHz images of the BAT AGN with jet-like morphologies. Three of these jetted objects (MCG+08-11-011, MCG-01-40-001, and NGC~5728) are relatively radio-bright, so the contours occur at 6$\sigma$, 12$\sigma$, and 18$\sigma$ above the background. In MCG+04-22-042 and MCG-02-08-014, additional contours are shown at 3$\sigma$. Each figure includes the beam (green, bottom left) and a scale bar representing 1 kiloparsec. The objects are shown at full 1\arcsec~resolution except for MCG+04-22-042 and MCG+08-11-011, which are shown at 3\arcsec~and 2\arcsec~resolution, respectively, to best illustrate the extent of their low surface brightness emission.}
    \label{fig:jetmorphs}
\end{figure*}


\subsection{Radio Data Reduction and Image Processing}
\label{sec:radioprocessing}

We conducted our observations in the K-band with the JVLA in the C-array configuration, resulting in 1\arcsec~spatial resolution. The K-band is centered at 22~GHz with a wide 8~GHz bandwidth. Observing blocks with 1-hour duration were shared among 2-3 targets, with each block beginning with X- and K-band attenuation scans and flux and bandpass calibrations with 3C~48, 3C~138, 3C~286, or 3C~147 depending on sky position and antenna wraps. Each science observation included between 3 and 10 minutes on-source integration time, based on time constraints from calibration overhead, and was preceded and followed by a gain calibration scan of a nearby source. The typical $1\sigma$~sensitivity in these observations is $\sim16\mu$Jy per beam. In addition to these initial observations, our most recent proposal round included deeper observations of 11 targets from the first campaign that we suspected had extended radio emission from star formation below our previous sensitivity limits \citep{Smith2016}; these images typically have $1\sigma$~sensitivities of $\sim8\mu$Jy. These 11 targets are denoted by asterisks in Table~\ref{t:tab1}, and have the same observing setup except that each object had between 13 and 20 minutes of on-source time. Four of these showed significant extended emission in the deeper images, all star formation-like. Since this analysis focuses on the origin of the core radio emission and the nature of the nuclear source, the results and implications for star formation are being analyzed for another paper in preparation. 

The reduction techniques are identical to those described in detail in \citet{Smith2016} for the initial sample. After collection, the raw data were passed through the standard JVLA reduction pipeline at the National Radio Astronomy Observatory (NRAO). We then processed the data using the Common Astronomy Software Applications package \citep[v. 4.5, CASA; ][]{McMullin2007}. Each individual object was split from the parent measurement set and averaged over all 64 channels within each spectral window in order to reduce processing time without compromising image quality. Each image was cleaned to a 0.03 mJy threshold using the CASA \texttt{clean} task with Briggs weighting, and then assessed for signs of pervasive and persistent RFI. In some cases, RFI affected only isolated spectral windows and could be corrected by flagging and removing the affected window. In other cases (those enumerated in Section~\ref{sec:radiosample}), the RFI was too widespread for effective removal and objects were discarded. If an image was bright enough (peak flux $\geq$ 1~mJy), we performed phase self-calibration of the visibility data. 

\subsection{Core and Extended Radio Flux Measurements}
\label{sec:radiofluxmeasurement}

For compact sources and the cores of extended sources, we use the CASA command \texttt{imfit} to fit each compact Stokes I image component with an elliptical gaussian. All of the compact cores and unresolved sources were well-fit by this technique. To explore the possibility of low surface-brightness emission, images were also made with 3\arcsec~and 6\arcsec~beam tapers. To measure the total extended emission in each object, we re-cleaned each image with a 6\arcsec~beam taper. We then use \texttt{imfit} on this larger-scale image. For structures extending beyond the 6\arcsec~beam, we measure the total flux manually in CASA \texttt{viewer} by using \texttt{imstat} on a custom elliptical region drawn around the emission; such a fit was required for 10 objects.

It is important to note that even though the core is compact in our 1\arcsec~resolution images, there still may be star formation or jet structure within the beam, convolved with the true AGN core component.

\section{Radio Morphologies}
\label{sec:morph}
We divide our images into four broad morphological categories: unresolved, resolved emission indicative of star formation (i.e., extended but non-linear), jet-like resolved emission (i.e., extended and linear), and objects that are compact at 22~GHz but have resolved jets at lower frequencies in archival observations. An unresolved core is present in all 96 radio-detected objects, including those with extended star formation. The total numbers of each type are 55/96 compact (four of which have jets at lower frequencies), 30/96 star formation, and 11/96 jets. Resolved objects from the original sample of 70 are shown in \citet{Smith2016}; objects with non-compact morphologies from the new sample are shown in Figure~\ref{fig:sfmorphs} (star formation) and Figure~\ref{fig:jetmorphs} (jets). Note that NGC~3227 is quite compact, however, we classify it as star formation-like due to the asymmetry of the resolved emission. 

\begin{figure*}
\begin{tabular}{cc}

      \includegraphics[width=0.5\textwidth]{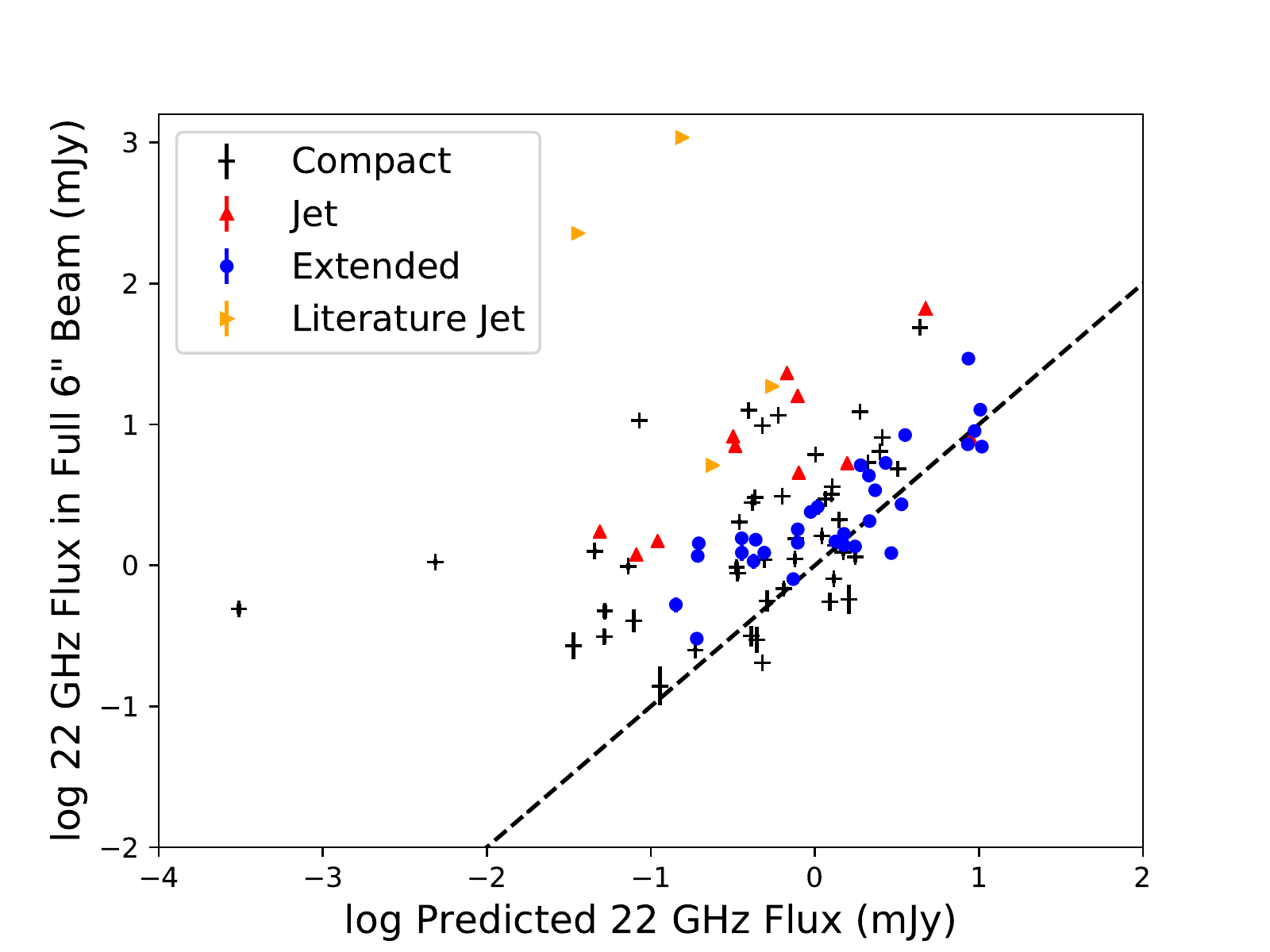} &
    
      \includegraphics[width=0.5\textwidth]{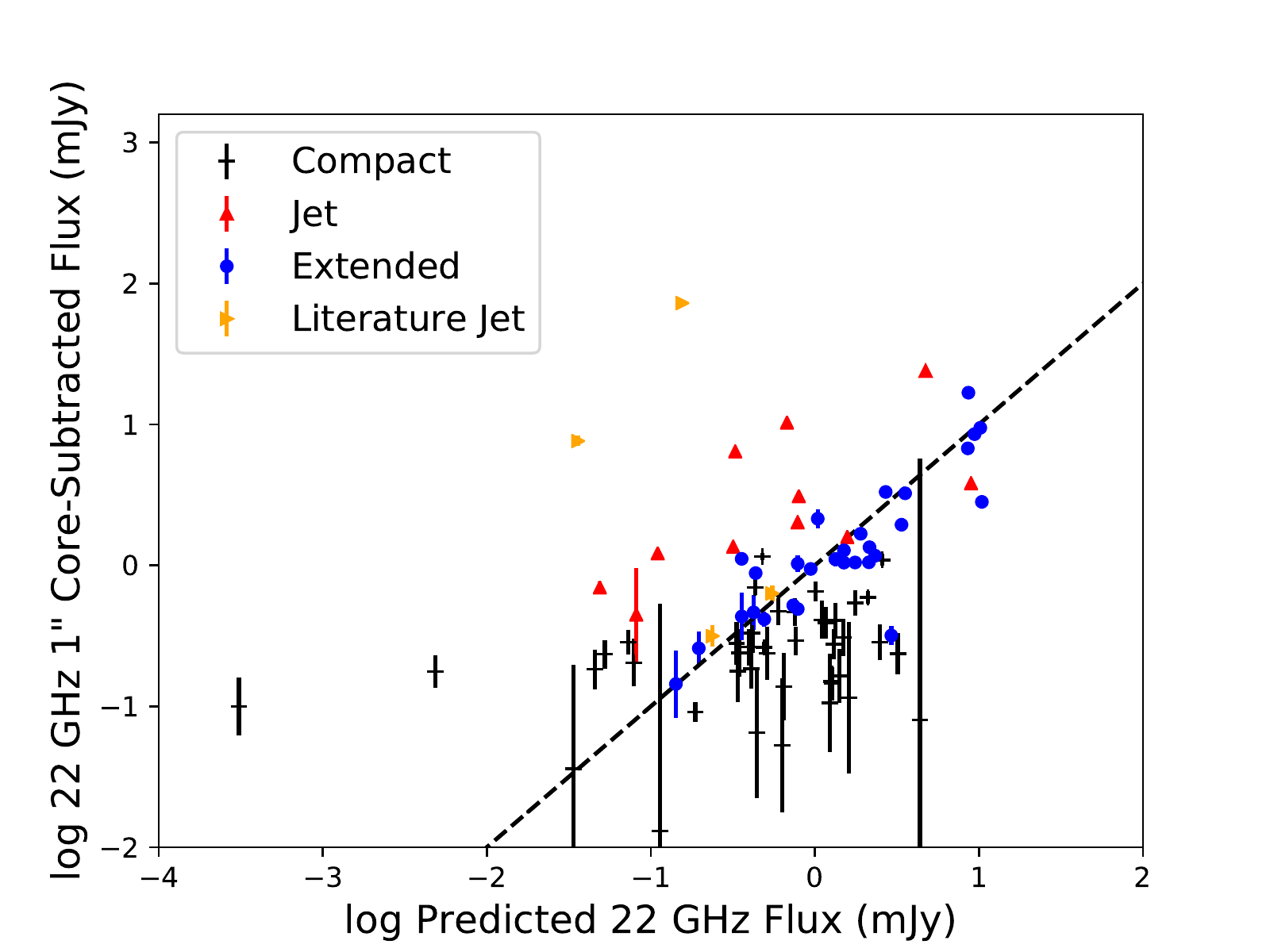} \\
    
      \includegraphics[width=0.5\textwidth]{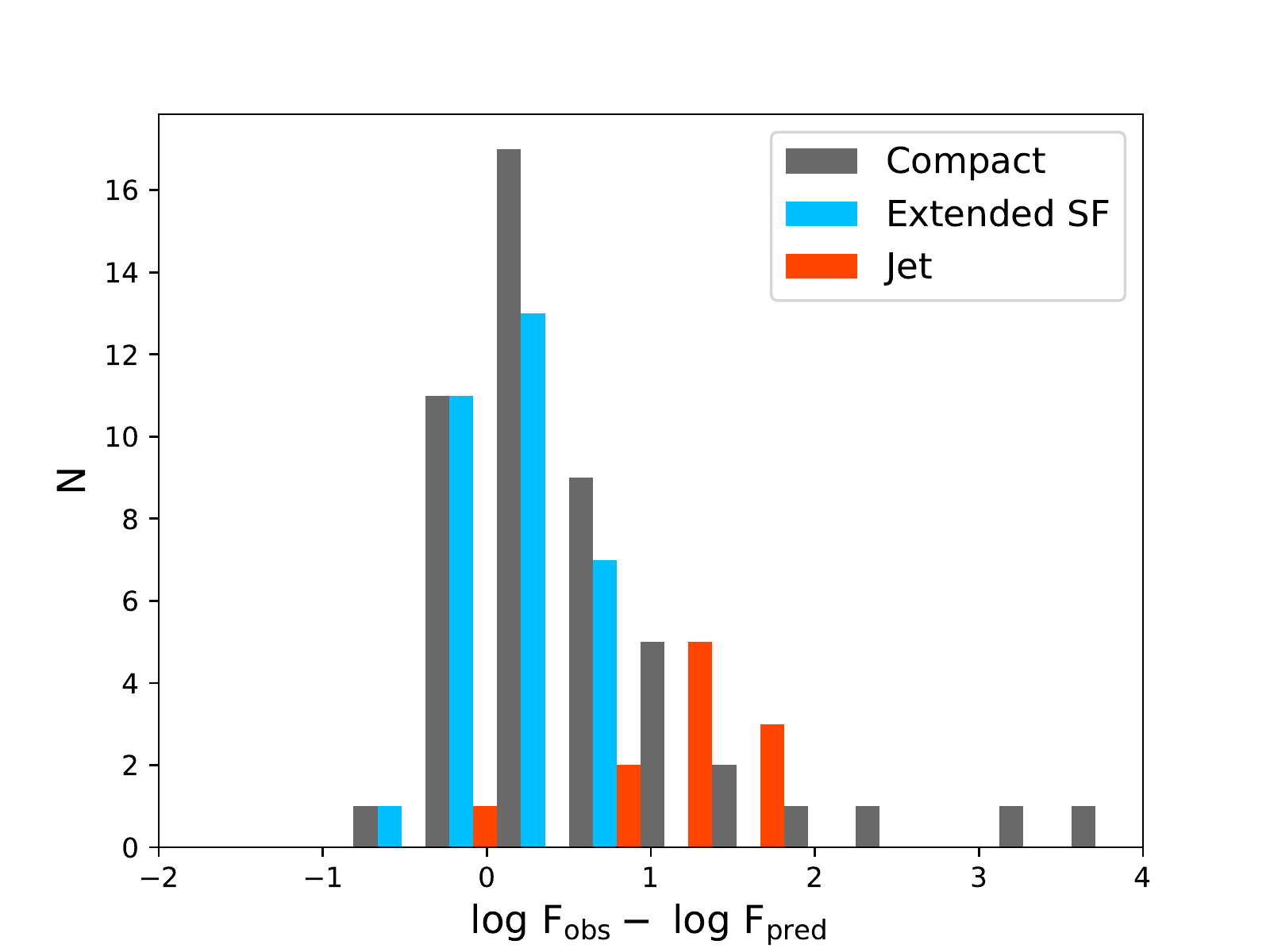} &
    
      \includegraphics[width=0.5\textwidth]{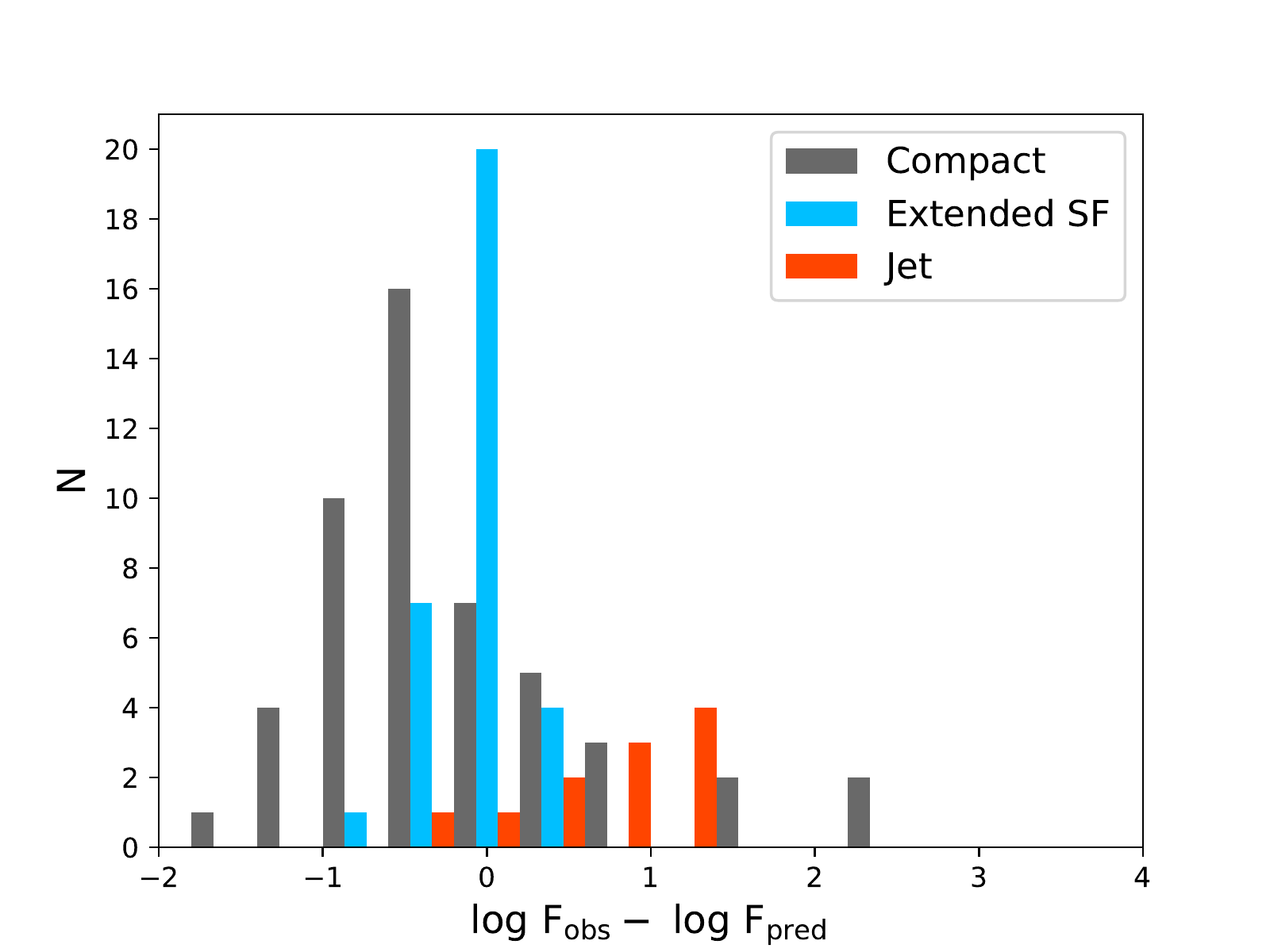}\\
      
\end{tabular}
\caption{Upper panels: the observed 22 GHz flux density versus the predicted flux density from star formation based on the infrared star formation emission for the full 6\arcsec~ beam, including all extended emission (left) and for the extended emission only, with the unresolved 1\arcsec~ core subtracted (right). Vertical error bars are shown for all points, but are often smaller than the point markers. Large error bars on unresolved sources after core subtraction result from the very low flux remaining when the unresolved core (obviously the majority of the flux in these cases) is removed. The dashed lines indicate the 1-to-1 relation between predicted and observed flux. Lower panels: histograms of the offsets of the measured fluxes from the 1-to-1 line. The literature jet sources are excluded due to their very low numbers and high offsets, to keep the horizontal range illustrative for the other classes.}

\label{fig:obsvpred}
\end{figure*}

To determine how much of the unresolved emission is due to star formation, we can compare the radio emission from \emph{only} the extended star formation, subtracting the unresolved core, to the far-infrared emission from star formation. If they match, then the subtracted core emission must be AGN related. \citet{Shimizu2017} presented far-IR spectral decomposition of the full \emph{Herschel} sample, quantifying the amount of FIR emission due to star formation and AGN respectively. With the SF-related infrared emission in hand, we can use the $L_\mathrm{FIR} / L_\mathrm{R}$ relation \citep{Condon1992} to calculate the expected radio emission from star formation. We then compare this expectation to the \emph{extended}, core-subtracted radio emission in Figure~\ref{fig:obsvpred}. The lower panel of the figure shows a histogram of the offsets of the measured extended fluxes from the star formation prediction. After core subtraction, the objects with extended star formation emission tend to match the star formation prediction from the infrared quite well, implying that the core emission is mainly AGN-related and the extended emission is indeed from star formation. As can be seen in the lower panel histogram, the extended flux in objects with star formation morphologies match the predicted values with low dispersion, $\sigma = 0.25$. Objects with jet morphologies remain systematically above the expectation for star formation after core subtraction, as expected if the extended radio emission is due primarily to jet synchrotron emission.

Note that many of the compact sources fall well below the star-formation expectation after core subtraction. This is possibly because some radio-emitting star formation is unresolved in the core. Star formation at all scales contributes to the infrared emission, and therefore the expected radio flux derived from it. However, any radio emission from star formation unresolved at 1\arcsec~  is removed by the radio core subtraction, causing the core-subtracted flux to fall below the FIR/radio expectation. The errors are large for these core-subtracted compact sources, since the majority of the flux has been removed in the subtraction.

Finally, we note that the morphologies of the jet-like objects are not sufficient by themselves to indicate a jet instead of star formation. As in \citet{Smith2016}, we also consider the ratio of the total observed radio emission to that predicted from the FIR-measured star formation rate, based on the $L_\mathrm{FIR} / L_\mathrm{R}$ relation \citep{Condon1992}. All of the jet-like objects except NGC~5728 exhibit radio emission a factor of $10-20$ times stronger than expected from the star-formation related infrared emission in our \emph{Herschel} observations, indicating the presence of an AGN-related jet. After subtracting the AGN core, NGC~5728 moves below the predicted line and NGC~3516 falls almost directly upon it (Figure~\ref{fig:obsvpred}.) This would be sufficient reasoning to reclassify these targets as likely to be star formation dominated, except that their radio emission has been well-studied and established as jet-like in the case of NGC~3516 \citep[e.g.,][]{Wrobel1988,Veilleux1993} and a combination of a radio jet and a star formation ring \citep{Durre2018} in NGC~5728, a structure also apparent in our image (Figure~\ref{fig:jetmorphs}). Because our conclusions focus on the core emission and whether it differs in objects with or without jets, not whether there is or isn't circumnuclear star formation, we leave this combination object as a jet in our figures and analyses.

Concerning the objects with compact 22~GHz morphologies but low-frequency jets: two of these sources were observed at higher resolutions than our survey, and the elongated structures are small enough to lie within our 1\arcsec~beam: MCG-01-24-012 \citep{Schmitt2001} and Arp~102B \citep{Helmboldt2007}. The other two had linear structures considerably larger: $\sim37$~kpc for NGC~3718 \citep{Condon1987} and 2.8~kpc for NGC~1052 \citep{Wrobel1984}. They may have been missed by us because optically-thin synchrotron emission in jets are typically steep-spectrum, and so are not easily seen at 22~GHz, or because energy losses may steepen the spectrum sufficiently above 1.4~GHz to prevent detection at our flux limit. Since we have checked for structure with multiple beam tapers out to 6\arcsec, it is unlikely that the jets are simply ``resolved out", although we may still be insensitive to larger structures seen in much lower resolution surveys like NVSS. Since not all of our targets had sufficiently high-resolution archival observations, we do not know whether radio jets exist at lower frequencies for all of our compact sources - for this reason, we refer to these objects as ``Literature Jets" and give these four objects a distinct symbol in the plots and tables so that the reader may treat them separately or consider them with the compact sources. Two of these are radio-loud objects Arp~102B and NGC~1052.

\begin{figure*}
\begin{tabular}{cc}

      \includegraphics[width=0.5\textwidth]{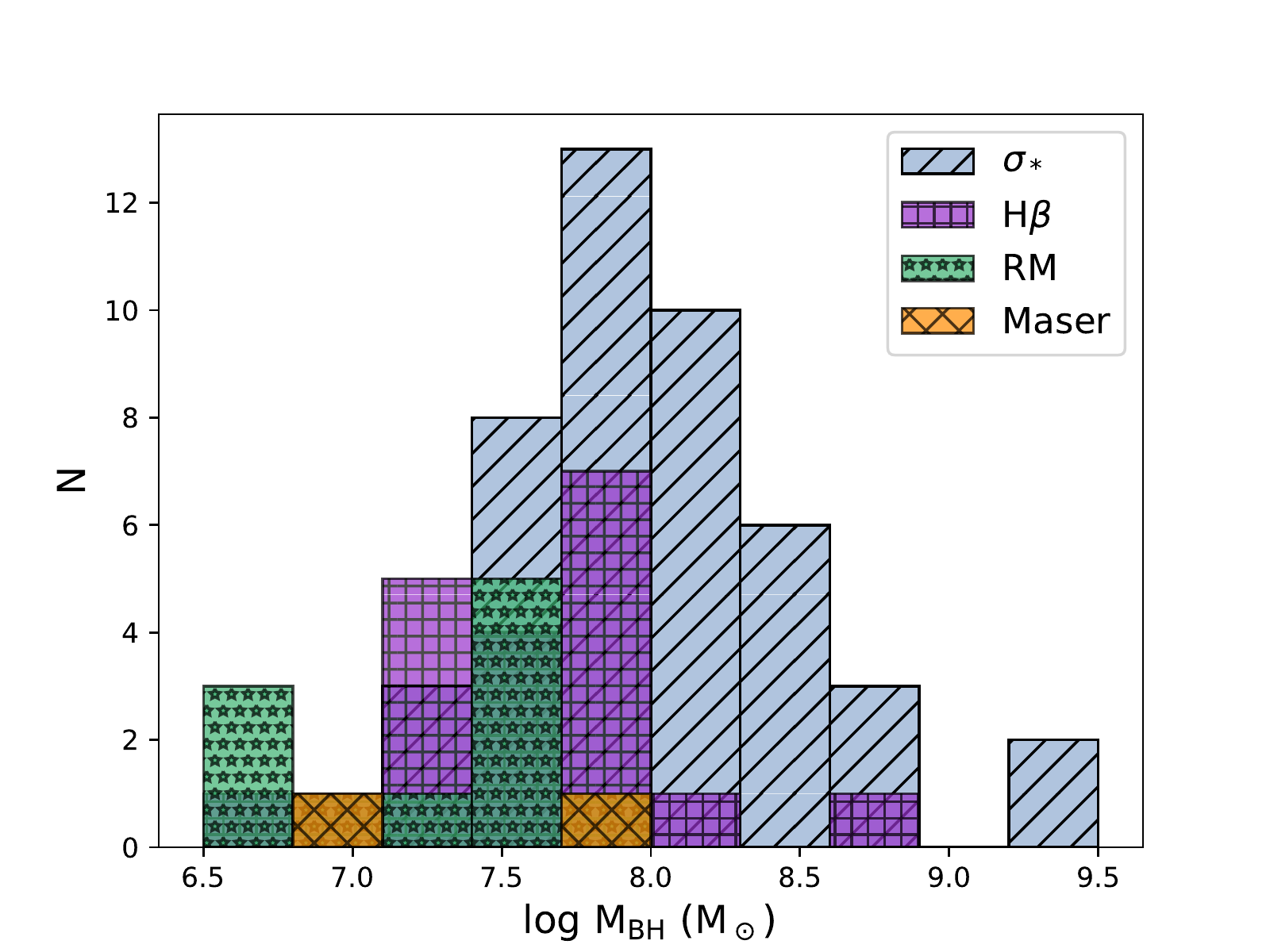} &
    
      \includegraphics[width=0.5\textwidth]{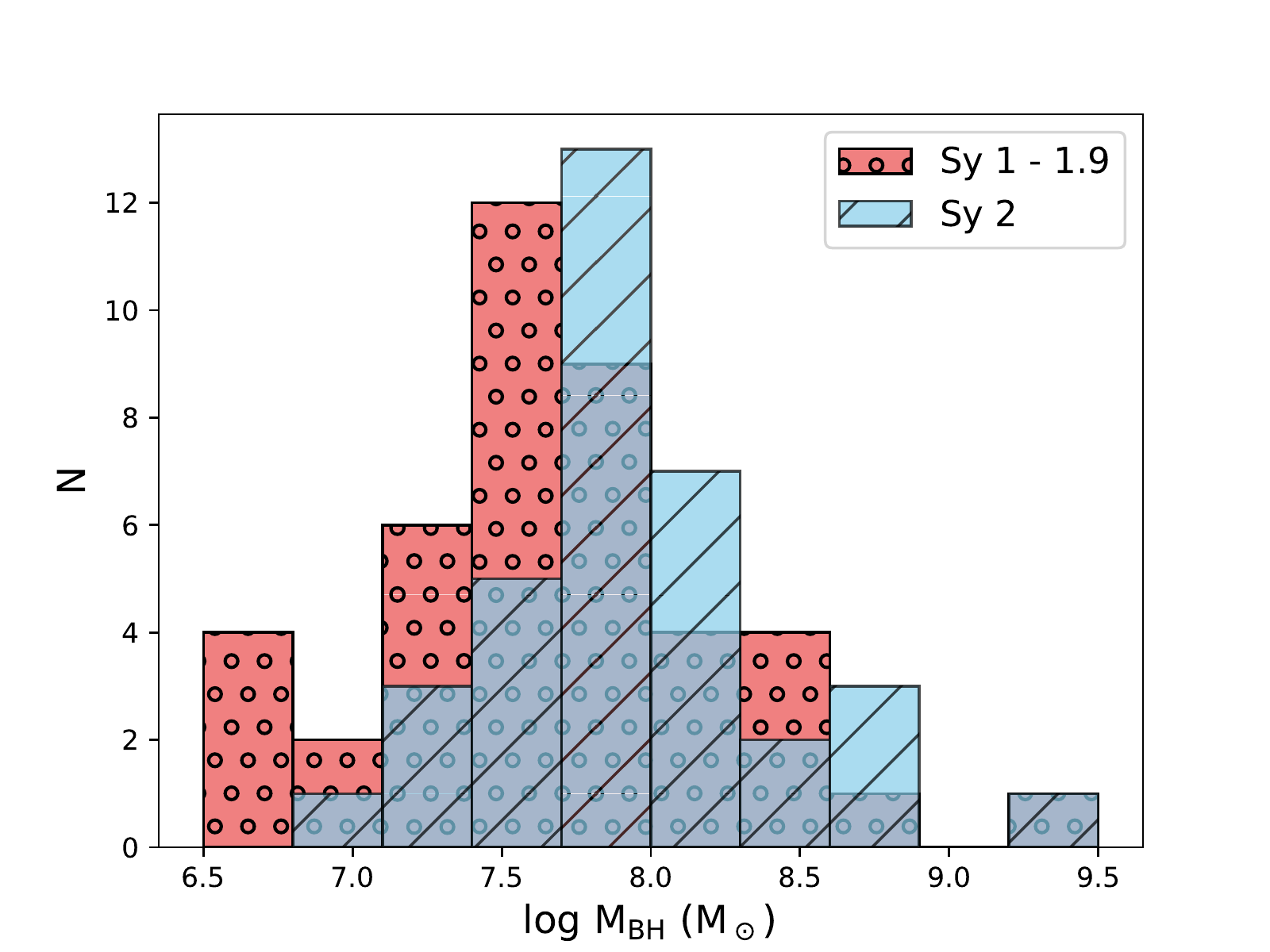} \\

      \includegraphics[width=0.5\textwidth]{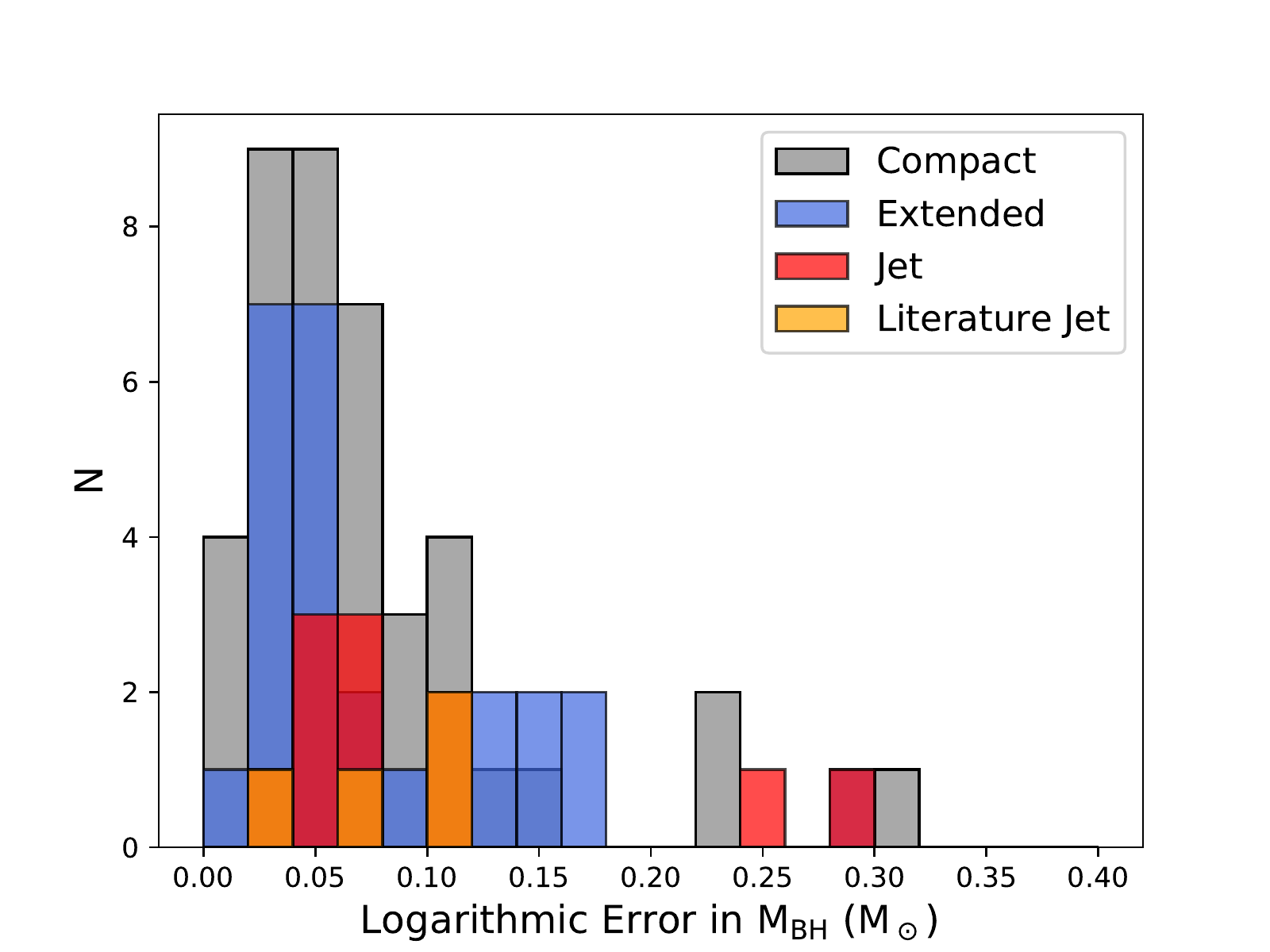} &
    
      \includegraphics[width=0.5\textwidth]{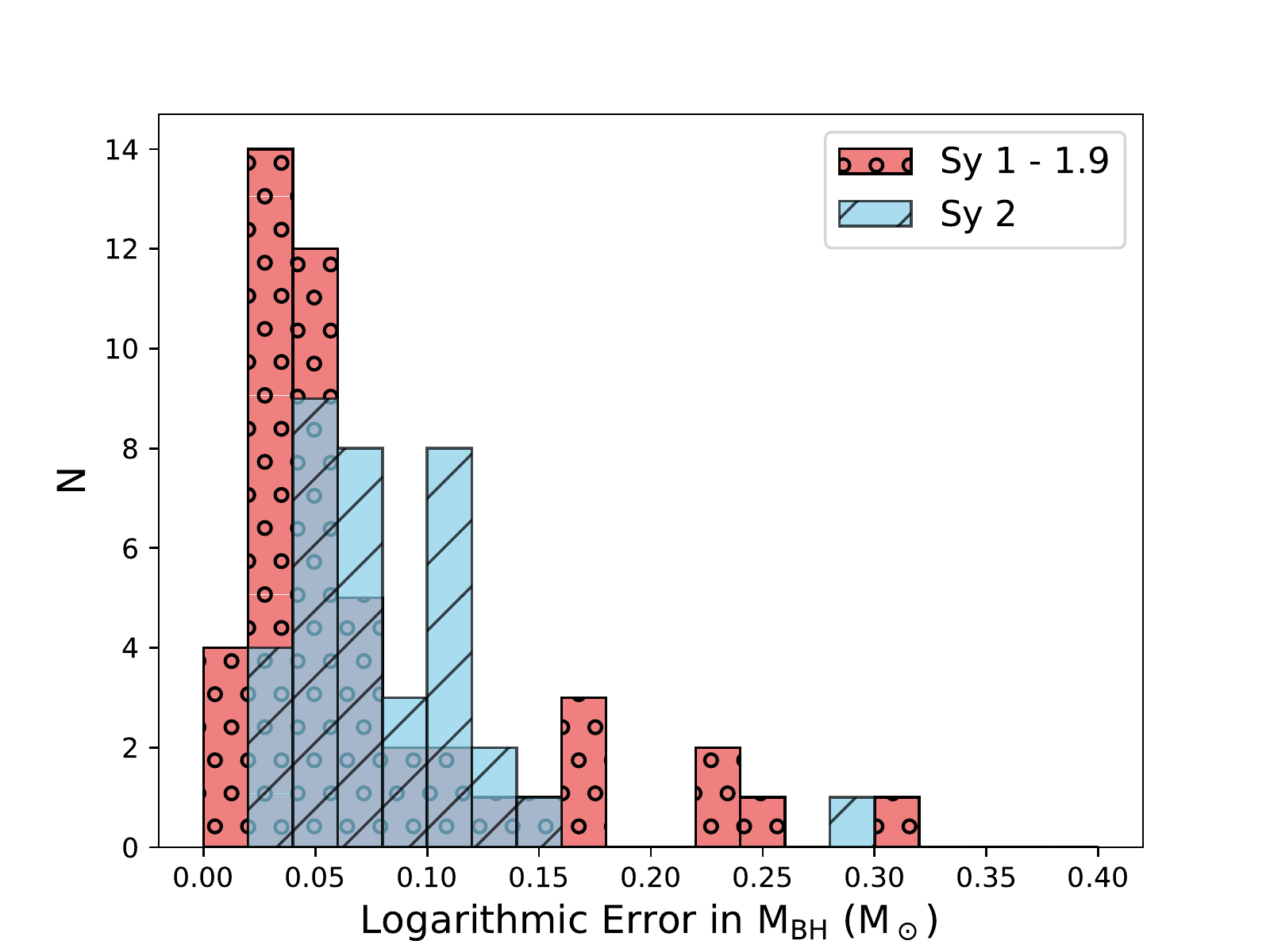} \\

\end{tabular}
\caption{Histograms of the black hole mass estimates generated by the four different measurement techniques (top left), black hole masses by Seyfert type (top right), black hole mass errors by 22~GHz radio morphological class (bottom left), and black hole mass errors by Seyfert type (bottom right).}

\label{fig:mbherrors}
\end{figure*}


\begin{figure*}
    \centering
    \includegraphics[width=\textwidth]{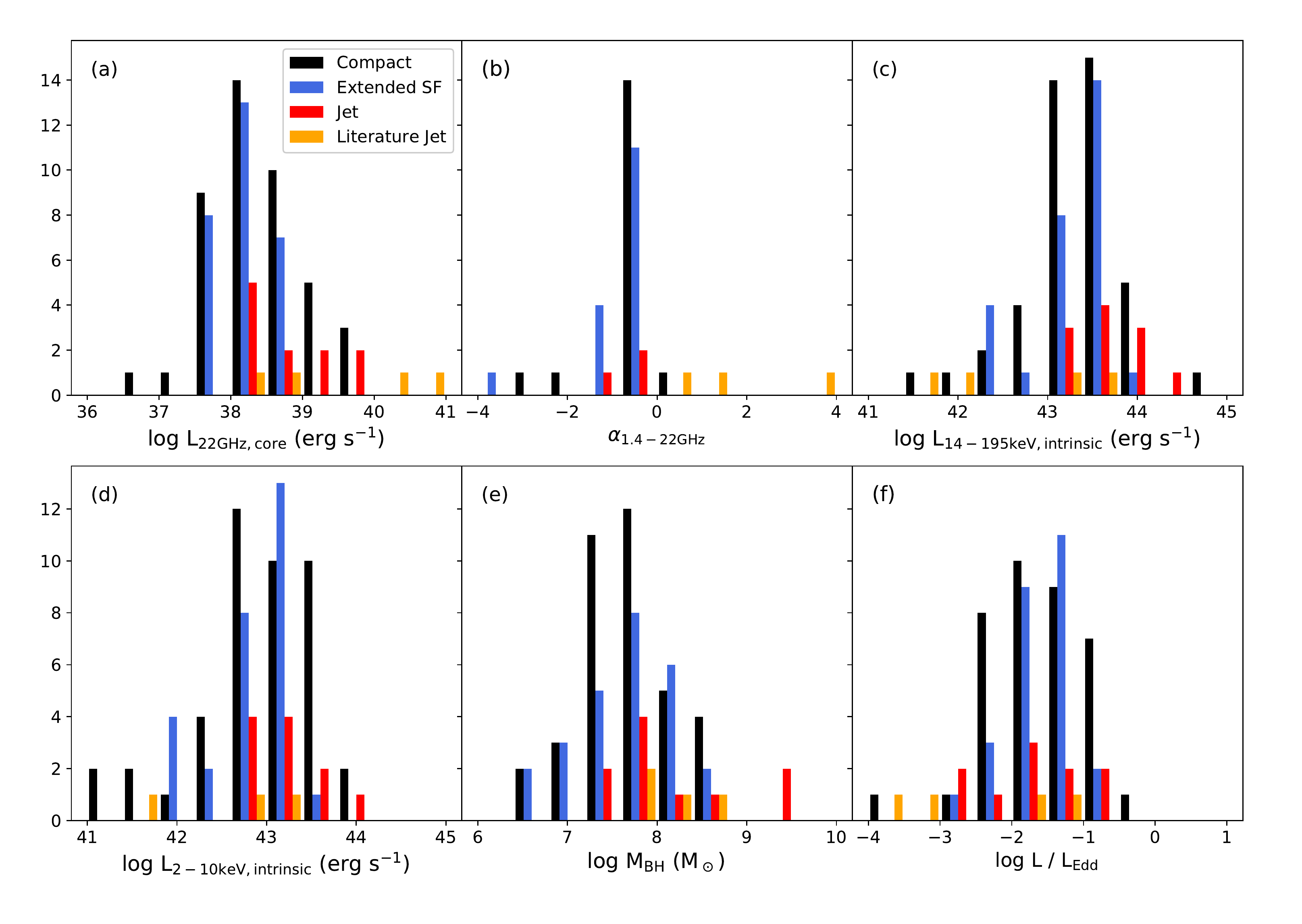}
    \caption{Histograms of (a) the 22~GHz radio luminosities inside the 1\arcsec~core, (b) the radio spectral indices between the 1.4~GHz FIRST flux and the 22~GHz 6\arcsec~ total flux, (c) absorption-corrected ultra-hard X-ray luminosities, (d) absorption-corrected hard X-ray luminosities, (e) black hole masses, and (f) Eddington ratios. Colors indicate different 22~GHz morphologies: compact or unresolved (black), extended star formation (blue), jets (red) and literature jets (orange). }
    \label{fig:allpar_histograms}
\end{figure*}


\section{Luminosity, Black Hole Mass, and Accretion Rate}
\label{sec:paramstats}

In this section we compare the X-ray luminosities, 22~GHz radio core luminosities, radio spectral indices, black hole mass estimates, and Eddington ratios for our morphological classes.

\subsection{X-ray Luminosities}
\label{sec:xraylums}
The observed X-ray luminosities provided in this paper were first reported by \citet{Baumgartner2013}, and were obtained by fitting a simple power law to the eight channels of the BAT instrument without accounting for obscuration. A more detailed treatment of the X-ray spectral properties of the BASS AGN is given by \citet{Ricci2017}. The X-ray spectra were fit with a large number of different models accounting for all components commonly observed in AGN, including photoelectric absorption and Compton scattering, which enables measurement of the absorption-corrected AGN flux. For the most obscured, Compton-thick ($N_{\rm H}\geq 10^{24}\rm\,cm^{-2}$) AGN we used a physical torus model to correctly constrain the line-of-sight column density and the intrinsic flux \citep{Ricci2015}. The result is a confident estimate of the hard ($2-10$~keV) and ultra-hard ($14-195$~keV) intrinsic X-ray luminosities. The dominant source of measurement error depends on the obscuration of the source; for highly obscured sources, errors on the intrinsic luminosity arise primarily from uncertainties in the fitting of the spectral absorption. For the majority of sources with low or moderate obscuration (N$_\mathrm{H} < 10^{23}$ cm$^{-2}$), the errors come from simple errors in the BAT spectra from which the fitted band luminosities were derived.
\\

Due to the nature of the more complex models used by \citet{Ricci2017}, in a few unobscured sources the reported intrinsic flux values are very slightly \emph{above} the ``observed" values. This is a reflection of an intrinsic uncertainty in the fluxes determined from a simple fit to the BAT data alone, as done by \citet{Baumgartner2013} to obtain the observed values: the median uncertainty in the X-ray photon index is $\Delta\Gamma = 0.15$ for the full BAT AGN sample. Sometimes, therefore, the ``observed" values in Table~\ref{t:tab1} will be slightly above the ``intrinsic" values derived from higher S/N joint fit of the whole X-ray spectral range including absorption, scattering, and a cutoff.

\subsection{Black Hole Masses}
\label{sec:mbh_calc}

The BASS survey DR2 has black hole mass estimates for 82 of our 100 surveyed AGN, using four different black hole mass measurement methods. Black hole mass estimates in AGN are subject to many uncertainties; by far the most reliable method is dynamical measurement of gas or stars under the direct gravitational influence of the black hole. However, this is impossible for all but the nearest AGN, none of which belong to our sample. Water maser emission is perhaps the next most reliable method, followed by reverberation mapping; both methods require fortuitous alignment and/or extensive monitoring campaigns, and so are still relatively rare \citep[e.g., ][]{Kormendy2013}. In our sample, two objects have maser mass estimates and 11 have been reverberation-mapped; note that reverberation mapping estimates are biased towards lower mass black holes, since the relevant timescales are shorter and easier to capture. The next best method uses the width of the broad emission lines to estimate the Doppler broadening due to motion of the gas within the gravitational influence of the black hole. In our sample, 22 masses were estimated from the width of the broad H$\beta$~emission line via the relation in \citet{Trakhtenbrot2012}. Finally, in the absence of broad emission lines one can use the stellar bulge velocity dispersion, $\sigma_*$, as a mass proxy through the M$_\mathrm{BH} - \sigma_*$~relation via \citet{Kormendy2013}, which was required for 47 of our objects. Mass estimates and associated errors are given in Table~\ref{t:tab1}. 

We note that the reliability of M$_\mathrm{BH} - \sigma_*$ mass estimates in AGN remains disputed: AGN may follow a different M$_\mathrm{BH} - \sigma_*$~relation than normal galaxies. The degree to which AGN adhere to M$_\mathrm{BH} - \sigma_*$ may depend on AGN properties: \citet{Dasyra2007} find that bright QSOs lie along a different relation than the generic Seyfert population, and \citet{Sheinis2017} find that radio-loud quasars follow a different relation than quiescent galaxies and radio-quiet AGN. However, \citet{Woo2013} and \citet{Shankar2019} find consistency between AGN and quiescent galaxies after accounting for selection effects. In any case, the M$_\mathrm{BH} - \sigma_*$~ relation is the only way to evaluate black hole mass or Eddington ratio for a sample that includes obscured AGN, and is more fundamental than the scaling relation between black hole mass and bulge luminosity \citep{Bernardi2007}.

At the time that \citet{Smith2016} was published, BASS data were not yet available. A few targets had $\sigma_*$~ measurements in the literature, typically from the 1990s \citep[e.g.,][]{Nelson1996}; these were excluded from the analysis given there. Since then, careful spectral measurements have become available from BASS with extensive comparison to the literature using improved spectral modeling codes with much better stellar templates \citep{Chen2014}. The measurements have benefitted from the inclusion of new spectra with much higher spectral resolution and better signal-to-noise ratios \citep[][Figure 18]{Koss2017}. The new spectra also have increased wavelength coverage and include additional stellar features that can be used to check and refine estimates of $\sigma_*$. As a result, the scatter in $\sigma_*$-derived mass estimates has reduced considerably within the BASS sample, and we are confident in the accuracy of those presented here. Note, however, that there is significant intrinsic scatter (0.3-0.5 dex) in the M$_\mathrm{BH} - \sigma_*$ relation \citep{Gultekin2009a}. This source of error in M$_\mathrm{BH}$ is more significant than any measurement errors of $\sigma_*$.

In Figure~\ref{fig:mbherrors}, we provide several histograms to assess the impact of various black hole mass estimate methods and their associated errors on the results presented later in this work. First, note that each measurement method has differing associated systematics; in our sample, masses measured using velocity dispersion tend to be higher than the other three methods. This results in Type~2 AGN, for which the velocity dispersion method is most common, having slightly higher black hole masses than Type~1 AGN in our sample (Figure~\ref{fig:mbherrors}, bottom right). A two-sided Kolmogorov-Smirnov test results in a 2.6\% chance that the distributions were drawn from the same sample. Note, however, that it is not yet established that Type~1 and Type~2 AGN have identical $M_\mathrm{BH}$ distributions, so offsets may be a combination of measurement methodology variations and intrinsic differences. Furthermore, the intrinsic scatter in M$_\mathrm{BH} - \sigma_*$ (0.3-0.5 dex)  is sufficient to render the offset between Type~1 and Type~2 AGN in Figure~\ref{fig:mbherrors} statistically insignificant.

Many of the plots later in this work compare quantities, some of which include $M_\mathrm{BH}$ in their calculation, between morphological subtypes. It is therefore important to know whether there are systematic differences in the measurement errors that could have arisen from, for example, a particular $M_\mathrm{BH}$  estimation method being more prevalent among objects in one of the radio morphological subtypes. No such systematic offsets are found after comparing lognormal fits of the subsample distributions (Figure~\ref{fig:mbherrors}, bottom left).

\begin{figure*}
\begin{tabular}{cc}

      \includegraphics[width=0.5\textwidth]{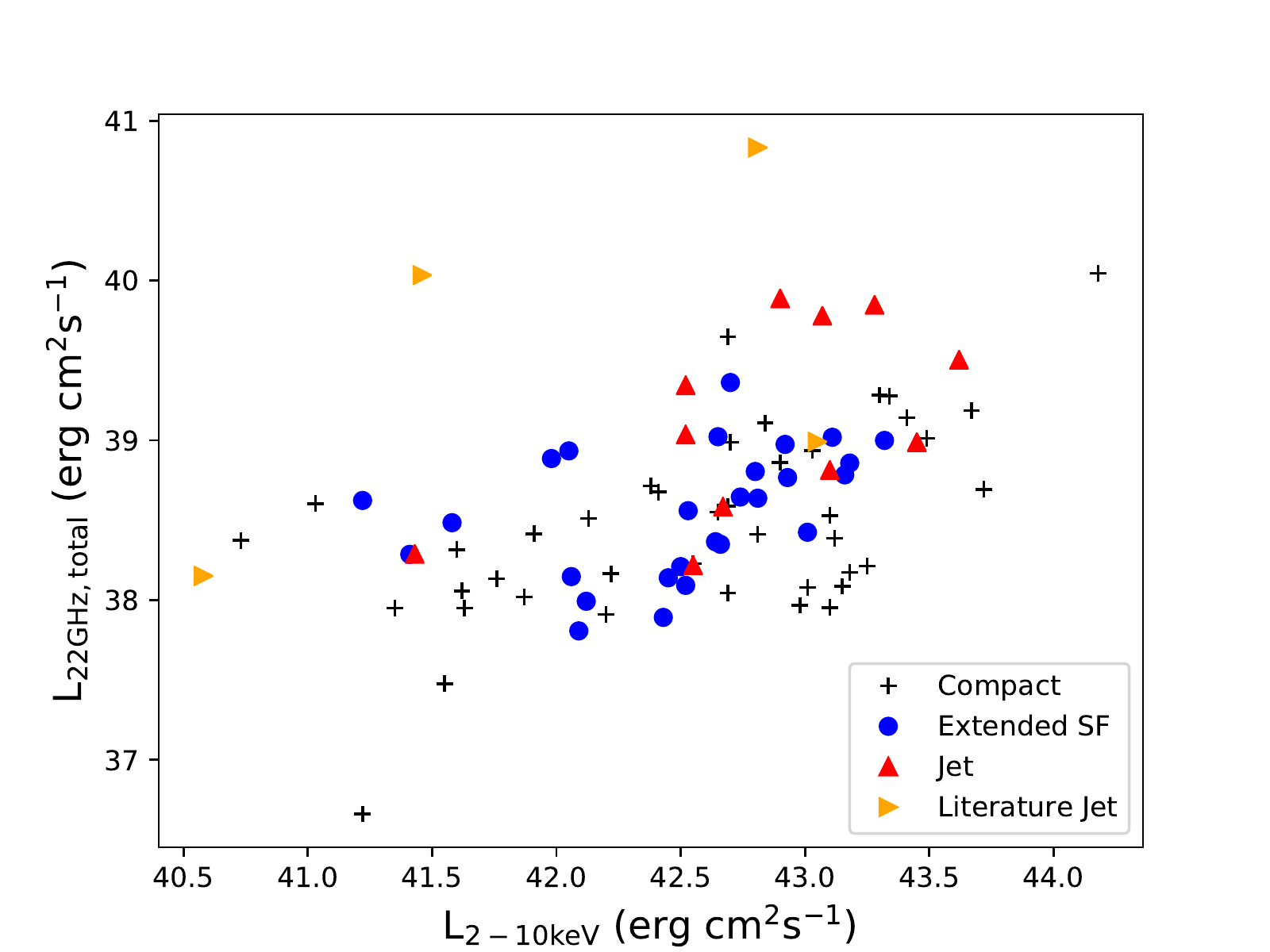}
      \label{fig:ceoc} &
    
      \includegraphics[width=0.5\textwidth]{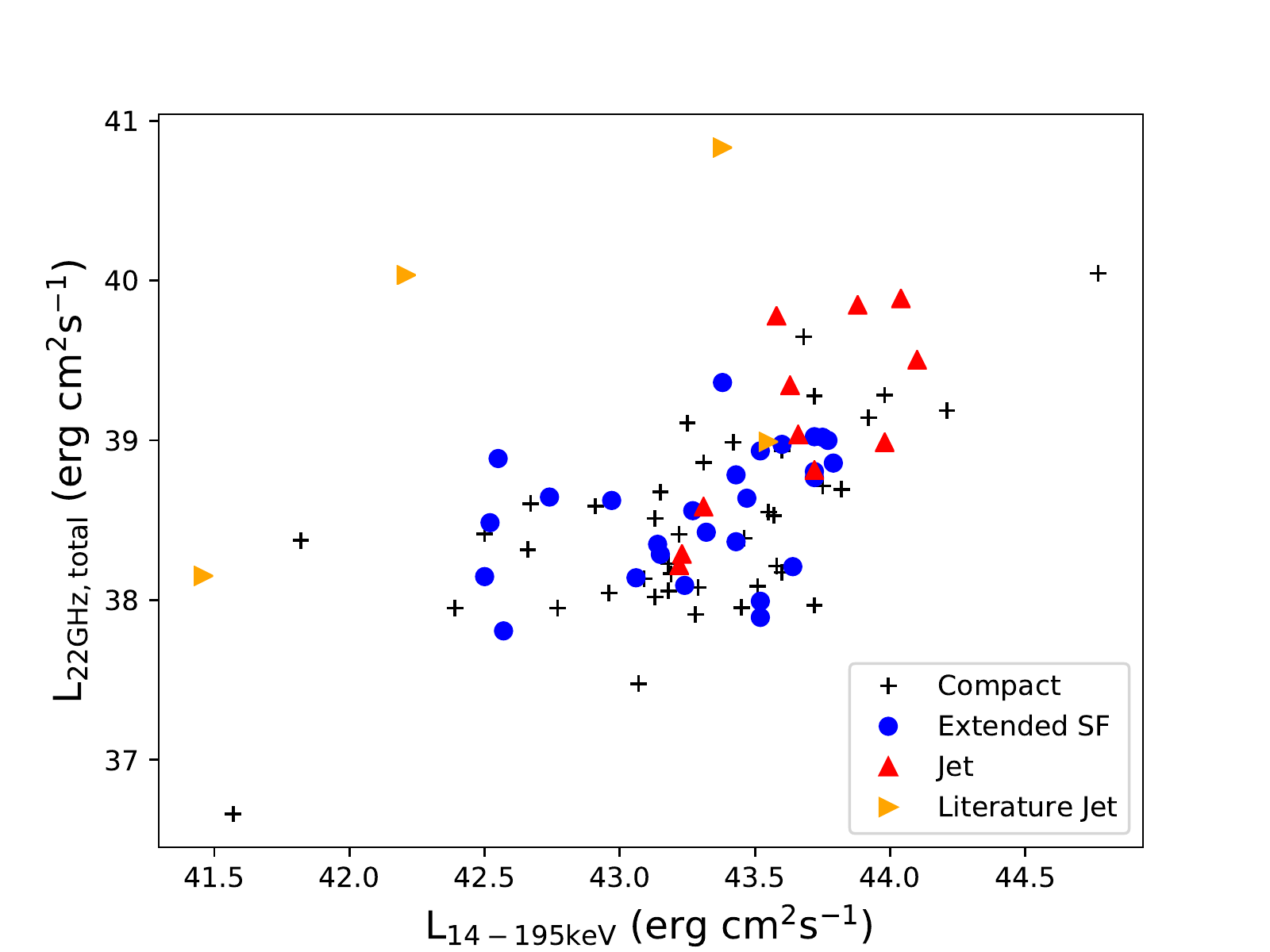}
      \label{fig:ceoc} \\

\end{tabular}
\caption{Total radio luminosity versus observed hard X-ray (left) and ultra-hard X-ray (right) luminosity. Different colors and symbols denote different 22~GHz radio morphologies as indicated in the legend.}

\label{fig:lrlx_morphdist}
\end{figure*}


For 18 of our objects, there are currently no black hole mass estimates, either because the object is not yet included in BASS, because the object was a pure Type~2 Seyfert with insufficient stellar absorption lines to estimate $\sigma_*$, or because the spectral region surrounding the H$\beta$~line was too noisy for a reliable broad line fit.

\subsection{Eddington Ratios}
We use $L_\mathrm{14-195~keV}$ and a bolometric correction factor to calculate the Eddington ratio, $L_\mathrm{bol}  / L_\mathrm{Edd}$. \citet{Vasudevan2009} found that for the BAT AGN sample, $L_\mathrm{bol} / L_\mathrm{14-195~keV} \sim 8$ as a median bolometric correction factor, which we adopt here; see also Section~3 of \citet{Koss2017}. We calculate the Eddington luminosity via $L_\mathrm{Edd}~= ~1.26\times10^{38} (M_\mathrm{BH} / M_\odot)$~erg~s$^{-1}$, as for all objects in BASS Data Release 1 \citep{Koss2017}.

\subsection{Radio Spectral Indices}
To compute the radio spectral index $\alpha$, we use archival 1.4~GHz fluxes from the FIRST survey \citep{Becker1995}, which includes 41 of our targets (mainly due to declination limits). Note when reviewing the spectral indices that the resolution of FIRST is $\sim5$\arcsec, compared to 1\arcsec~for our radio cores. For this reason, we calculate the spectral index using the 6\arcsec~fluxes (Section~\ref{sec:radiofluxmeasurement}), which is a closer match to the FIRST beam. Therefore, by necessity, some non-AGN emission is included in the spectral index measurement for objects with extended circumnuclear star formation. The 5\arcsec~ resolution is the primary reason behind choosing FIRST over the more complete NVSS \citep{Condon1998}, which has a much larger beam of $\sim$45\arcsec. Although approximately 1/3 of our sample has 5~GHz flux measurements from various sources in the literature, the beam size and sensitivity vary enormously, so we do not include these values in our calculation. 

Histograms of all the physical parameters for our sample compared by 22~GHz morphology are shown in Figure~\ref{fig:allpar_histograms}.

\section{$L_R/L_X$ Correlation}
\label{sec:lrlx}

We here investigate the relationship of the X-ray and radio emission in two ways. First, we determine whether or not the total radio luminosity (including extended emission) versus X-ray luminosity can be used to distinguish the dominant radio morphology. Second, we focus on the X-ray and radio properties of only the unresolved core to test whether any core properties are more likely to give rise to the observed extended morphologies.

\begin{figure}
    \centering
    \includegraphics[width=\textwidth]{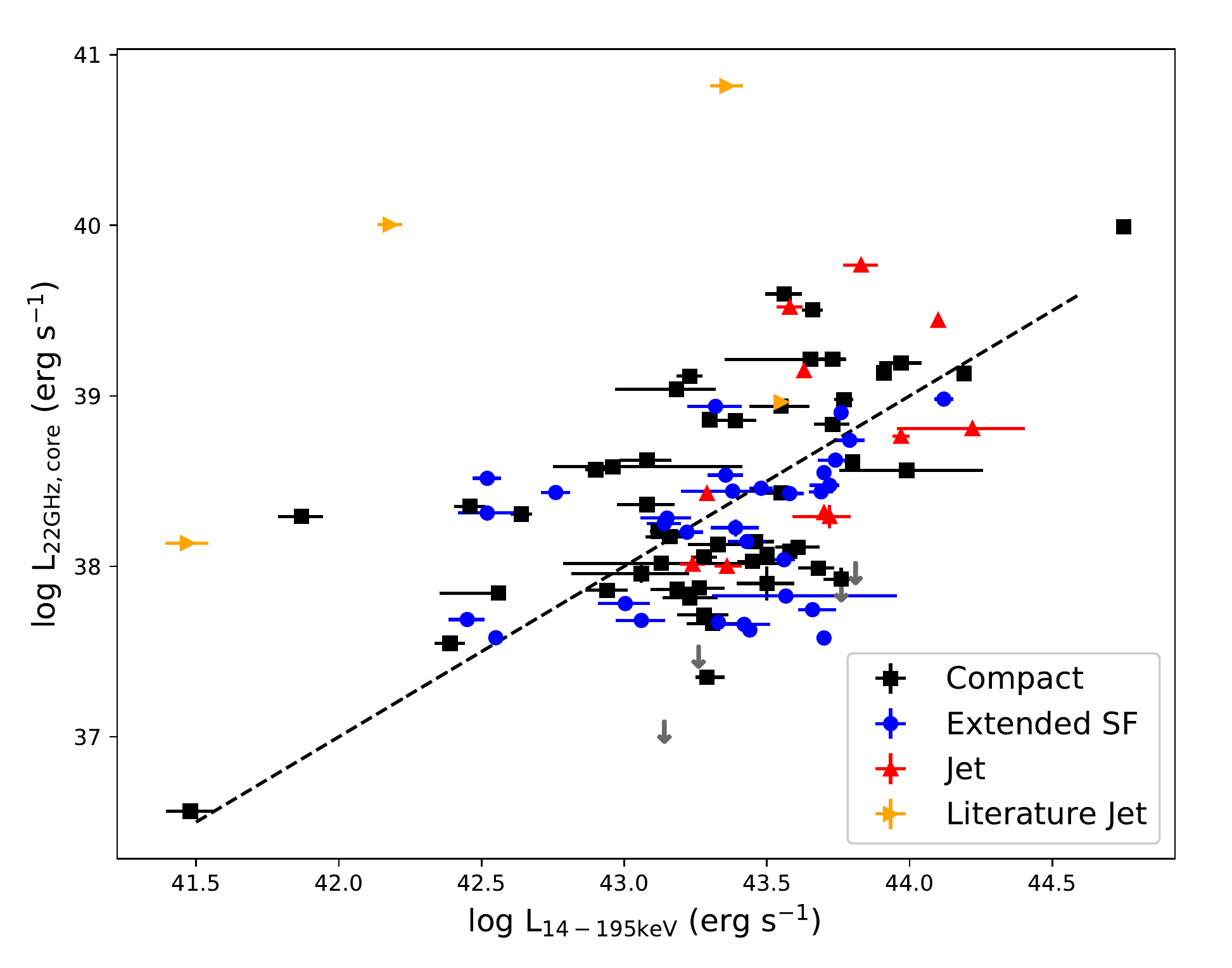}
    \caption{The 22~GHz luminosity in only the unresolved radio core versus the intrinsic, absorption-corrected ultra-hard X-ray luminosity. Error bars (1-sigma) are shown in both dimensions but are smaller than the data points for the radio luminosity. Errors in X-ray luminosity are primarily due to uncertainties in the absorption correction fitting for obscured sources and measurement errors in the X-ray spectrum for low-obscuration sources (Section~\ref{sec:xraylums}). The dashed line shows the $L_R/L_X$~ relation for stellar coronae from \citet{Guedel1993}.}  
        \label{fig:lrlx}
\end{figure}

Figure~\ref{fig:lrlx_morphdist} compares the total 22~GHz radio luminosity (combined core and extended components), to the hard and ultra-hard X-ray luminosities for the different 22~GHz morphologies in our sample. There is no combination of luminosities above which jets are more likely to dominate the morphology, or below which one finds prominent nuclear star formation.

In Figure~\ref{fig:lrlx}, we plot the relationship between the core 22~GHz radio and ultra-hard X-ray luminosities and fluxes. Recall that all targets have an unresolved radio core, regardless of the extended morphology; it is the emission from the core only that is plotted here. To control for the hidden mutual dependence on redshift, we calculate the partial Kendall's-$\tau$ correlation coefficient using the method of \citet{Akritas1996}. Table~\ref{t:pcor} shows the results of the partial correlation analysis for the full sample and the morphological subsamples. After the redshift is accounted for, the full sample and the compact sample separately still exhibit a significant correlation between radio and X-ray luminosity; objects with extended and jet morphology do not have significant $L_R/L_X$ correlations after distance considerations. Although the correlation coefficient is significant in some cases, the errors in the intrinsic X-ray luminosities, which are largely due to uncertainties in accounting for obscuration in the spectral fitting (Section~\ref{sec:xraylums}), preclude a determination of a precise correlation in our sample. We note, however, that these X-ray fits are state-of-the-art for obscured sources, especially given that the obscured BAT sample represent the brightest obscured AGN on the sky \citep{Ricci2017}. So, these errors reflect our current best ability to measure the true nuclear X-ray luminosity in the presence of significant intrinsic obscuration. Characterizing the $L_R/L_X$~ relationship for obscured AGN remains a work in progress.

We overplot the $L_R/L_X$~ relation for stellar coronae from \citet{Guedel1993}, and find that our objects are broadly consistent with this expectation, just as \citet{Laor2008} found for their sample of radio-quiet AGN. Predictably, the only significant outliers are the objects with known radio jets seen in the literature at lower frequencies, which are almost certain to have a significant contribution to their unresolved core flux from jet components. 

We note that the \citet{Guedel1993} relationship was established using X-ray data from a number of different telescopes at varying energy ranges, and that the \citet{Laor2008} application to AGN was performed at 5~GHz and 0.2-20~keV. Because research into the expectations for pure coronal emission in AGN at high radio frequencies and ultra-hard X-rays is still nascent, we refrain from attempting to predict how $L_R / L_X$~ may change. Very recent work by \citet{Behar2018} has established that the empirical relation is  $L_R / L_X \sim 10^{-4.2}$~ for very high radio frequencies in the mm-band (100~GHz) and at the BAT energies (14-195~keV), so there is some evidence for flattening at very high frequencies, but we do not know if this sets in as early as 22~GHz, and if the flattening is linear the effect at 22~GHz would be minimal, resulting in an expected value very near $L_R / L_X \sim 10^{-5}$.

Note that the four objects without radio detections, shown as upper limits (grey arrows) in the luminosity panel of Figure~\ref{fig:lrlx}, are not anomalously faint in the radio given their X-ray luminosities; it is therefore quite possible that they have radio cores within the normal range and slightly below our sensitivity limits.

Some of the scatter in the relationship is possibly due to the lack of simultaneity of our observations. \citet{Soldi2014} analyzed the \emph{Swift}-BAT light curves of AGN and found that Seyfert galaxies exhibited significant variability on timescales of months to years, although the fact that our luminosities are given as averages over many years of observations should mitigate the variability effect. At 22~GHz the variations are far slower, on the order of several years \citep{Hovatta2007}; however, these values are for radio-loud AGN and blazars, which are known to be much more variable than Seyferts in other wavebands. \citet{Baldi2015} found that the radio-quiet AGN in NGC~7469 varied by $\sim30$\% over 70~days at 95~GHz, and \citet{Mundell2009} found evidence for 8.4~GHz variability across many years in eleven Seyfert galaxies. In general, though, little is known about the variability of radio-quiet AGN at high frequencies.

The large scatter in the luminosity relation may also indicate that in order to understand the true relationship between these two quantities, additional parameters need to be considered. One quantity that is very likely to contribute to both luminosities is the black hole mass, which leads us to the fundamental plane.

\begin{center}
\begin{table}
\begin{tabular}{lcccccc}

\hline
\multicolumn{7}{c}{Full Sample}\\
\hline
$X$	&	$Y$	&	$Z$	&	N	&	$\tau$	&	$\sigma$	&	P$_\mathrm{null}$	\\
\hline
$L_X$	&	$L_R$	&	$z$	&	96	&	0.219	&	0.0643	&	$6.59\times10^{-4}$	\\
$L_X$	&	$L_R$	&	$M_\mathrm{BH}$	&	87	&	0.298	&	0.0698	&	$1.96\times10^{-5}$	\\
$L_R$	&	$M_\mathrm{BH}$	&	$L_X$	&	87	&	0.2	&	0.0769	&	0.0091	\\
$L_X$	&	$M_\mathrm{BH}$	&	$L_R$	&	87	&	0.128	&	0.0765	&	0.0935	\\
\hline
\multicolumn{7}{c}{Compact}\\
\hline
$X$	&	$Y$	&	$Z$	&	N	&	$\tau$	&	$\sigma$	&	P$_\mathrm{null}$	\\
\hline
$L_X$	&	$L_R$	&	$z$	&	49	&	0.248	&   0.0832	&	0.0029	\\
$L_X$	&	$L_R$	&	$M_\mathrm{BH}$	&	37	&	0.33	&	0.0985	&	$8.07\times10^{-4}$	\\
$L_R$	&	$M_\mathrm{BH}$	&	$L_X$	&	37	&	0.0356	&	0.105	&	0.735	\\
$L_X$	&	$M_\mathrm{BH}$	&	$L_R$	&	37	&	0.147	&	0.122	&	0.228	\\
\hline
\multicolumn{7}{c}{Extended}\\
\hline
$X$	&	$Y$	&	$Z$	&	N	&	$\tau$	&	$\sigma$	&	P$_\mathrm{null}$	\\
\hline
$L_X$	&	$L_R$	&	$z$	&	31	&	0.0906	&	0.134	&	0.499	\\
$L_X$	&	$L_R$	&	$M_\mathrm{BH}$	&	26	&	0.171	&	0.148	&	0.25	\\
$L_R$	&	$M_\mathrm{BH}$	&	$L_X$	&	26	&	0.461	&	0.151	&	$2.26\times10^{-3}$	\\
$L_X$	&	$M_\mathrm{BH}$	&	$L_R$	&	26	&	0.204	&	0.138	&	0.139	\\
\hline
\multicolumn{7}{c}{Jet}\\
\hline
$X$	&	$Y$	&	$Z$	&	N	&	$\tau$	&	$\sigma$	&	P$_\mathrm{null}$	\\
\hline
$L_X$	&	$L_R$	&	$z$	&	11	&	0.202	&	0.15	&	0.178	\\
$L_X$	&	$L_R$	&	$M_\mathrm{BH}$	&	10	&	0.338	&	0.114	&	$3.02\times10^{-3}$	\\
$L_R$	&	$M_\mathrm{BH}$	&	$L_X$	&	10	&	0.126	&	0.166	&	0.448	\\
$L_X$	&	$M_\mathrm{BH}$	&	$L_R$	&	10	&	-0.0632	&	0.192	&	0.742	\\
\hline
\end{tabular}
\caption{Results of partial correlation analysis for X-ray luminosity, radio luminosity, redshift, and black hole mass. The first three columns list the independent ($X$), dependent ($Y$), and third influencing ($Z$) variable. The remaining columns are (4) N, the total number of objects in the sample or subsample, (5) the partial Kendall's $\tau$~ coefficient calculated using the method of \citet{Akritas1996}, (6) $\sigma$, the square root of the variance of $\tau$, and (7) the probability $P_\mathrm{null}$~ of accepting the null hypothesis that $X$ and $Y$ are uncorrelated once the influence of $Z$ is accounted for. }
\label{t:pcor}
\end{table}

\end{center}


\section{The Fundamental Plane of Black Hole Activity}
\label{sec:fundplane}

Since X-ray luminosity is frequently used as a proxy for accretion rate and radio luminosity as a proxy for jet power, it is perhaps natural to assume that each of these quantities would be related to the black hole mass, and that these three quantities together might be related by a scale-invariant structure for the resulting object: an accreting supermassive black hole, efficiently transforming the potential energy of its fuel into hard X-ray emission and producing a radio-emitting jet. \citet{Merloni2003} tested this hypothesis by constructing a ``fundamental plane of black hole activity" spanning from stellar-mass galactic black holes in various different accretion states to supermassive black holes in a wide variety of AGN types. The plane takes the form:

\begin{equation}
\mathrm{log} ~L_{R} = \xi_{\mathrm{RX}}~\mathrm{log}~L_{X} + \xi_{\mathrm{RM}}~\mathrm{log}~M + K
\end{equation}

\noindent \citet{Merloni2003} found values of $\xi_{\mathrm{RX}} = 0.60$, $\xi_{\mathrm{RM}} = 0.78$ and $K=7.33$ for the coefficients and constant. In order to test the fundamental plane for only AGN targets, excluding stellar mass black holes, \citet{Bonchi2013} compiled a sample of hard ($2-10$~keV) X-ray selected AGN of both optical Seyfert types with 1.4~GHz radio observations at 1\arcsec~resolution, intentionally excluding the extended radio emission as we do. This sample is therefore quite similar to ours in many ways. The coefficients that they found for the fundamental plane (Equation~1) are $\xi_{\mathrm{RX}} = 0.39$, $\xi_{\mathrm{RM}} = 0.68$ and $K=16.61$.

To determine whether black hole mass is significantly correlated to the X-ray or radio luminosity once the $L_R/L_X$ dependence is removed, we perform the same partial correlation analysis as for the luminosities and distances, and provide the results in Table~\ref{t:pcor}. When considering the full sample, we find that the radio luminosity and black hole mass are likely to be correlated after the removal of the influence of $L_X$, but that the X-ray luminosity is not related to the black hole mass. The same result was found in the partial correlation analysis by \citet{Merloni2003}.
 
Since the ultra-hard X-rays and high radio frequencies in our sample are potentially well-suited for studying the AGN core properties (see Section~\ref{sec:corona_discussion}), a fundamental plane constructed from these parameters may offer new insight.

Using the black hole mass estimates from the BASS survey as described in Section~\ref{sec:paramstats}, we compare our targets' measurements to the fundamental planes found by \citet{Merloni2003} and the more similar sample of \citet{Bonchi2013}. In the discussion (Section~\ref{sec:discussion}), we consider the different selection criteria that can lead to discrepant results.  

None of the existing fundamental planes have been built using high-frequency radio observations, mainly because they tended to take advantage of existing large 1.4~GHz and 5~GHz surveys. Additionally, the X-rays used in these samples are in the $2-10$~keV rather than $14-195$~keV range of \emph{Swift}-BAT. We do have $2-10$~keV measurements for our targets from a variety of X-ray telescopes \citep[for details see ][]{Ricci2017}, both absorption-corrected and uncorrected. In all fundamental plane plots, we use the 22~GHz emission from the unresolved core as the observed radio quantity. The physical arguments underlying the fundamental plane do not indicate that star formation should be related to the X-ray emission or black hole mass, however there maybe other mechanisms which could cause a correlation between these parameters. It is important to note, however, that despite our small spatial beam extents of a few hundred parsecs it is still possible that the unresolved component includes some star formation emission.

\begin{figure*}
\centering
\begin{tabular}{ll}

      \includegraphics[width=0.43\textwidth]{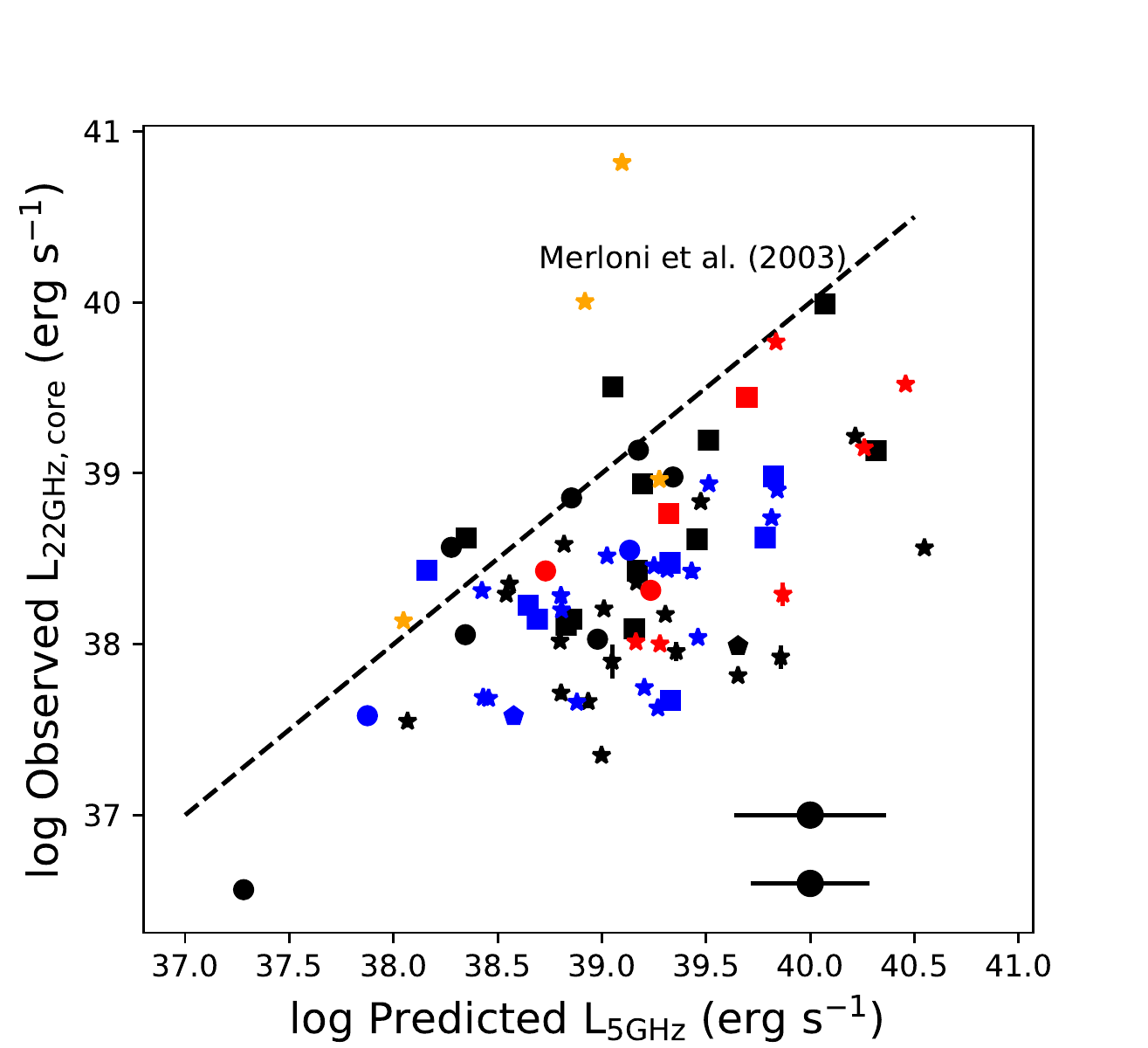}
 &
    
      \includegraphics[width=0.43\textwidth]{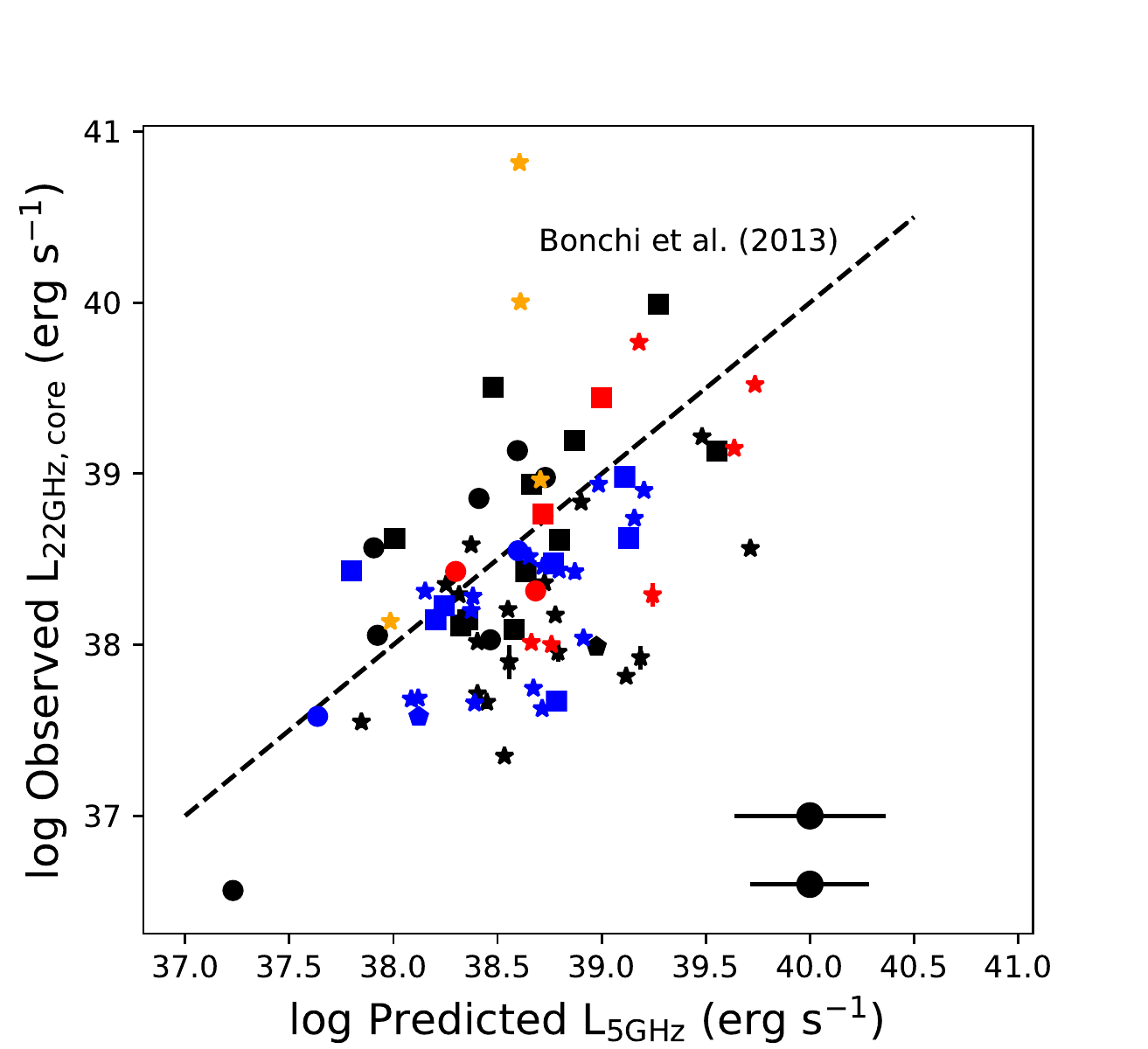}
 \\
 
       \includegraphics[width=0.5\textwidth]{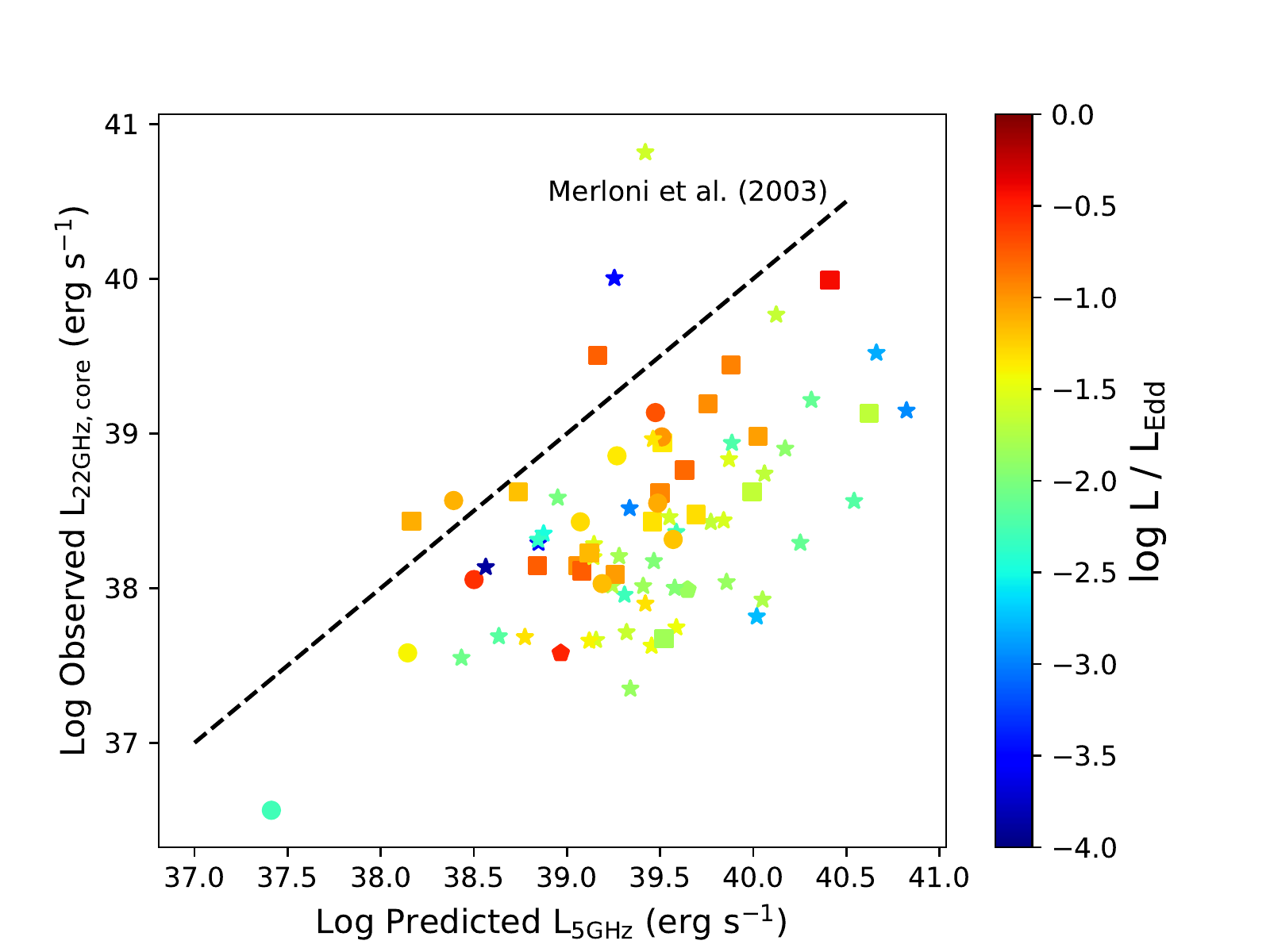}
 &
    
      \includegraphics[width=0.5\textwidth]{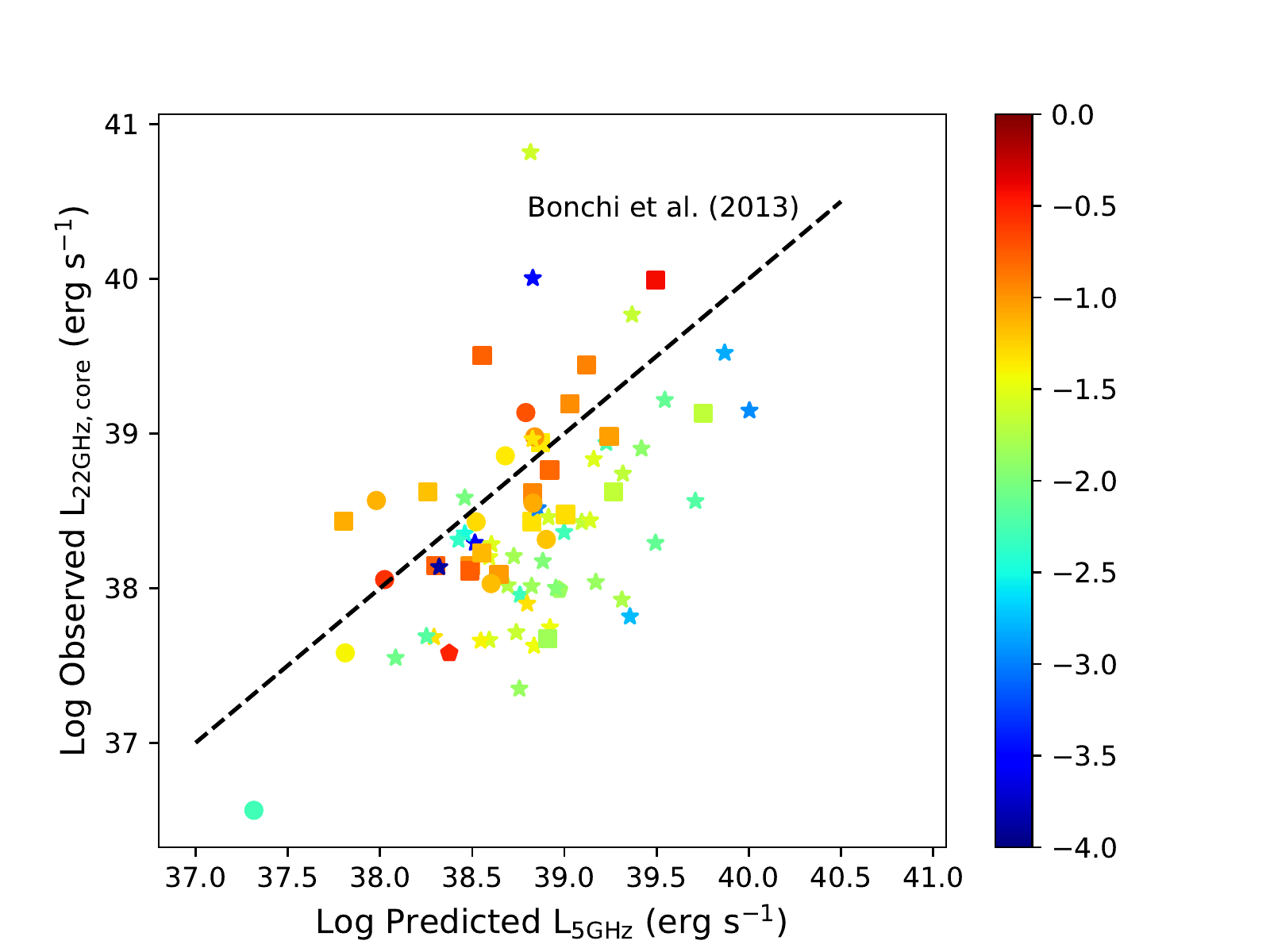}
 \\

\end{tabular}
\caption{Comparison of the BAT AGN sample's core radio flux to the \citet{Merloni2003} and \citet{Bonchi2013} fundamental plane predictions (based on the intrinsic $2-10$~keV luminosities). The dashed line is the 1:1 relation. Top: Colors denote the four categories of 22~GHz radio morphology (Section~\ref{sec:morph}): compact (black), extended star formation (blue), jet-like (red), and compact at 22~GHz but with jets at other frequencies in the literature (orange). Bottom: Colors indicate the log of the Eddington ratio as coded in the accompanying colorbars. In all figures, symbol shape denotes the $M_\mathrm{BH}$ measurement method (Section~\ref{sec:paramstats}): stellar velocity dispersion (star), H$\beta$~width (square), reverberation mapping (circle), and masers (pentagon). Errors on the observed radio luminosities are shown in upper plots but invisible for most points. Error bars in the lower right corner indicate the possible spread in the predicted luminosity due to X-ray absorption correction in the worst case (highly absorbed, larger bar) and best case (unabsorbed, smaller bar). Most of the width of the bar consists of the mean error in the black hole mass estimates.}
\label{fig:bonchi_merloni_planes}
\end{figure*}

Figure~\ref{fig:bonchi_merloni_planes} shows how our objects compare to the \citet{Merloni2003} original plane and the \citet{Bonchi2013} plane, which is more similar to our sample. Our targets lie systematically below the \citet{Merloni2003} fundamental plane (i.e., having less observed radio emission than predicted), just as the initial small sample did in \citet{Smith2016}. This remains true even when we transform the observed 22~GHz luminosities into inferred 5~GHz luminosities, the original quantity used by \citet{Merloni2003}. We do this in one of two ways: if the object has a 1.4~GHz detection in the FIRST survey, we interpolate between this value and our 22~GHz point; if not, we assume a radio spectral index of $\alpha=-0.7$. Our sample is much better matched to the \citet{Bonchi2013} fundamental plane. Because the validity of including objects into the fundamental plane may depend on accretion rate (see discussion in Section~\ref{sec:discussion}), we also include in Figure~\ref{fig:bonchi_merloni_planes} the same relations, but color-coded by Eddington ratio.

The dominant source of error in our measurements come from the black hole mass estimates, as shown by the typical error bars in the figures. Due to the vagaries of measuring black hole masses (see Section~\ref{sec:paramstats}), these are not atypical error bars for fundamental planes. Referring back to Figure~\ref{fig:mbherrors}, we note that there is no systematic difference in the magnitude of the errors between the radio morphological subtypes, so any differentiation in proximity to the planes by subtype is not due to a particular measurement method being more prevalent for that type. We note also that \citet{Gultekin2009b} constructed a fundamental plane using only high-quality, dynamically-measured masses with low error, but their sample is at much lower luminosity (and redshift) than ours, with very low accretion rates, and so is not especially comparable.

We also compare our observations using both X-ray absorption corrections and ultra-hard X-ray luminosities to the fundamental plane, and find that there is significant scatter and that objects lie well below the plane, regardless of nuclear radio morphology. This same conclusion was reached by \citet{Smith2016} with respect to the \citet{Merloni2003} fundamental plane; although that smaller sample had only 16 objects with mass measurements.

Our sample is more closely aligned to the \citet{Bonchi2013} fundamental plane, but the objects with star formation 22~GHz morphologies fall somewhat below that relation as well. 

If we broaden the picture to include stellar mass black holes our sample aligns well with predictions despite the scatter. \citet{Plotkin2012} considered a number of previous samples and reviewed the statistical difficulties in each one regarding the use of the planar coefficients to determine the X-ray emission mechanism. They created their fundamental plane using Bayesian regression and including stellar-mass black holes, Sagittarius A*, low-luminosity AGN, and BL Lacs. We show the \cite{Plotkin2012} fundamental plane along with our own sample in Figure~\ref{fig:plotkin}. The BAT AGN lie along the relation, but with greater scatter: the standard deviation of the distance between the observed points and the best-fit fundamental plane are 0.68 for our BAT AGN sample, 0.48 for the low-luminosity AGN from Plotkin, and 0.26 for the X-ray binaries from Plotkin.

\section{Discussion}
\label{sec:discussion}

\subsection{Radio Detection Fraction}
Of our sample of 100 ultra-hard X-ray selected AGN, only four sources were undetected at 22~GHz with the JVLA at our sensitivity level. The 96\% radio detection fraction for this ultra-hard X-ray selected sample with $-4 < \mathrm{log}~L / L_\mathrm{Edd} < 0.5$  is a strong argument against the existence of radio-silent AGN. In our sample, even when significant star formation emission is detected, there is always an unresolved radio core.

We note that this is a much higher detection fraction of BAT-selected AGN than found by \citet{Burlon2013} using 20~GHz data from the AT20G survey; however, their limiting flux density was 40~mJy, higher than all but 4 of our detections. Our detection fraction also contrasts with the results at 1.6-2.2~GHz by \citet{Roy1998}, who found that most of their sample of radio-quiet Seyfert galaxies lacked radio cores at a 5$\sigma$ sensitivity limit of 8~mJy and 3~mJy for each frequency, respectively. Either 22~GHz observations are more likely to recover radio cores in radio-quiet AGN despite being generally fainter than lower-frequency observations, or previous samples suffered from inadequate sensitivity or selection effects geared towards non-detections. The \citet{Roy1998} observations had a sensitivity limit of $\sim$3~mJy, but at a higher resolution of 0.1\arcsec. Our much higher detection fraction at lower sensitivity thresholds supports the hypothesis that sufficiently sensitive observations will recover radio cores in all radio-quiet AGN. 

However, attempts to detect radio cores in radio-quiet AGN using VLBA observations at 1.4~GHz with very high sensitivities do not find radio cores at anywhere near our observed occupation fraction. \citet{Maini2016} find that only 2 out of 4 radio-quiet AGN had a VLBI core, and \citet{Herrera-Ruiz2016} found that only 3/18 radio-quiet quasars in the COSMOS field had VLBI cores. Note, though, that these samples are at significantly higher redshifts. A nearby sample of Seyferts studied by \citet{Baldi2018}, however, successfully detected radio cores at 1.5~GHz in all 4 objects in their sample. 

If 22~GHz observations are actually more successful at finding radio cores in radio-quiet AGN, this may be a much more effective way to search for such cores than resource-intensive VLBI.

\subsection{Physical Parameters and Radio Morphologies}
\label{sec:discussion_physpar}

Figure~\ref{fig:allpar_histograms} compares the core radio luminosities, radio spectral indices, X-ray luminosities, black hole masses, and Eddington ratios for our four morphological subsamples. As expected, the jets have higher core radio luminosities than the other samples. Jets also have among the flatter spectral indices, and higher black hole masses, although the separation is not statistically significant.

Objects with compact and star formation morphologies reach to lower black hole masses than jetted objects by an order of magnitude. This may indicate that objects with higher black hole masses are more able to drive kiloparsec-scale jets than those with lower masses; however, one would expect that the ability to drive a jet would depend most crucially on accretion rate. Indeed, there are a number of theoretical implications that certain accretion disk geometries are more capable of launching and sustaining a jet than others \citep[e.g., ][]{Wiita1991, Blandford1999, Tchekhovskoy2015}. While the literature jets do indeed reach low values of $L / L_\mathrm{Edd}$, the distribution in Eddington ratio does not differ significantly between our morphological subsamples.

Finally, we note that objects with jets tend to have higher X-ray luminosities than objects with compact or star formation-dominated morphologies. 

It is possible that the jet's interaction with material along its propagation is generating X-ray emission in addition to that emitted by the corona, increasing the X-ray luminosities of these targets as seen in NGC~4258, although this may be a rare phenomenon \citep{Cecil1992,Yang2007}. It is also possible that the same electrons responsible for the radio synchrotron emission are emitting in the X-rays. The amount of X-ray luminosity expected from a jet compared to its radio luminosity varies widely across quasars, from $\sim0.3-10$\% \citep{Schwartz2010}, and is known only for large-scale jets, so it is difficult to know if this is consistent with expectations. In radio-loud and -intermediate quasars, the jet is indeed known to contribute to the X-ray luminosity \citep{Miller2011}. 

Note also that the jets' higher X-ray luminosities may be due to a somewhat subtle selection effect: at 22~GHz, thermal bremsstrahlung radiation begins to become comparable in importance to synchrotron radiation in star-forming galaxies \citep{Condon1992}. Therefore, the bremsstrahlung component may make star-forming regions easier to detect than jet lobes with observations of the same sensitivity. So, potentially, only relatively high-luminosity jets may be detectable, contributing to their somewhat higher luminosities in Figure~\ref{fig:allpar_histograms}.

\subsection{$L_\mathrm{R}/L_X$ and Radio Morphologies}
\label{sec:lrlx_discussion}

Figure~\ref{fig:lrlx_morphdist} compares the $L_\mathrm{R,total}/L_X$ relationship for different morphological subsamples.

In the right panel of Figure~\ref{fig:lrlx_morphdist}, where we plot the total radio emission (including resolved) against the ultra-hard X-ray, a trend is only clearly seen when the radio morphology is dominated by jets or the core. Objects with star formation morphologies contributing significantly to the radio emission do not correlate with the observed $14-195$~keV X-rays, as one would expect.

We also note that jet-like and star formation radio morphologies remain well mixed even at low values of the total radio luminosity. Since much of the structure in our radio maps would be unresolved by the 5\arcsec~ and larger beams of most large surveys (e.g., FIRST, NVSS), or even high-resolution surveys at higher redshifts, this is a cautionary note against using radio luminosity to determine the nature of the unresolved emission as star formation or AGN-powered.

\subsection{Coronal vs. Scaled-down Jets via $L_R/L_X$}
\label{sec:corona_discussion}

For decades, X-ray and radio emission have been observed to be tightly correlated in AGN samples with a wide range of bolometric luminosities \citep[e.g,][]{Brinkmann2000,Panessa2007}. X-ray emission is ubiquitous among AGN, and its rapid variability implies highly nuclear origins within a few gravitational radii of the black hole. The X-ray emission originates from UV and optical accretion disk photons that are Compton-upscattered by a population of electrons above the accretion disk in a hot, compact plasma known as the ``corona" \citep[e.g., ][]{Shapiro1976,Haardt1991,Wilkins2012}. The geometry and relationship between the corona and the disk is an active field of research. Such a structure comprised of electrons in a magnetic field would also necessarily emit in the radio; so, at least some radio emission in all AGN that have coronae must be due to this and not to scaled-down versions of relativistic jets. In radio-quiet AGN, the coronal component may dominate, whereas in radio-loud AGN it is overwhelmed by emission from the jets.

\begin{figure}
    \centering
    \includegraphics[width=\textwidth]{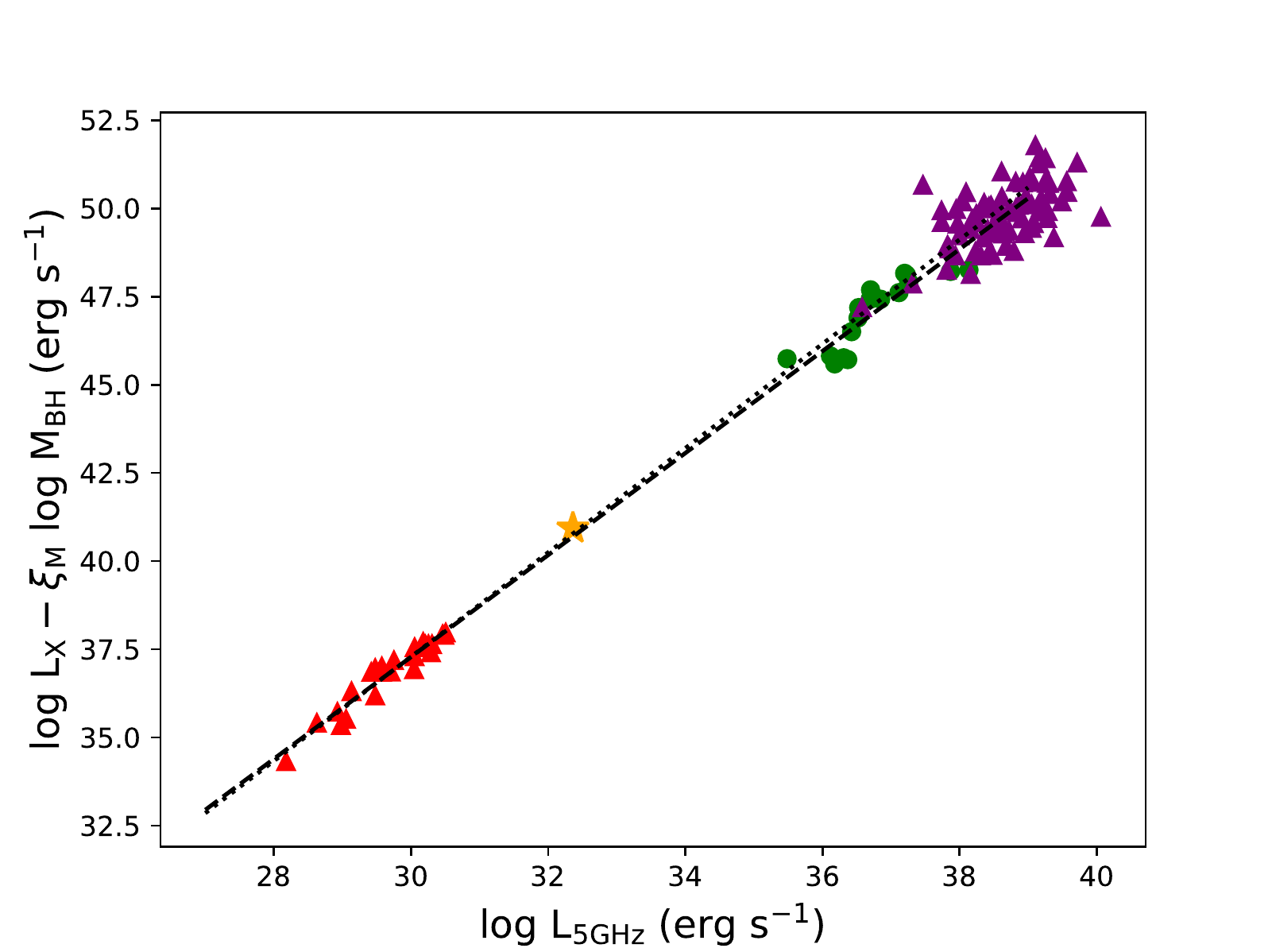}
    \caption{The fundamental plane relationship from \citet{Plotkin2012} including their sample of stellar mass black holes (red triangles), Sagittarius A* (gold star), and low-luminosity AGN (green circles). The present sample of BAT AGN is shown in purple. The dotted line is the best linear regressive fit to only the stellar mass black holes, and the dashed line is the best linear regressive fit to the entire Plotkin sample (including LLAGN and Sgr~A$^*$).}
        \label{fig:plotkin}
\end{figure}


If the origin of the bulk radio emission is significantly different between radio-loud and radio-quiet sources, one might expect the $L_R/L_X$ relationship to be different between them as well. \citet{Laor2008} showed that both populations exhibit highly significant correlations between 5~GHz radio luminosity and $0.2-20$~keV X-ray luminosity, but that the radio-loud sources were distinct, with values of $L_R\sim10^{3}$ higher than radio-quiet quasars with similar $L_X$. A similar result was found by \citet{Capetti2007}. Furthermore, \citet{Laor2008} postulated that the radio emission in their radio-quiet sample was due to coronae analogous to those found in stars, and found that their X-ray and radio luminosity relation was consistent with the $L_R/L_X \simeq 10^{-5}$ relation for cool, coronally active stars, found by \citet{Guedel1993} using archival ROSAT and \emph{Einstein} data and 5~GHz VLA observations. The relationship is quite tight and linear, and holds from the stars, through ultraluminous X-ray sources, and out to radio-quiet quasars. 

As discussed in Section~\ref{sec:lrlx} and shown in Figure~\ref{fig:lrlx}, the core radio and ultra-hard X-ray luminosities in our sample are consistent with the \citet{Guedel1993} relation. 

Our data set may be especially well-suited to studying the coronal properties of AGN: ultra-hard X-rays are not contaminated by star formation, coming uniquely from the AGN component, and are less affected by obscuration than hard X-rays. Additionally, our high-frequency 22~GHz observations may sample a small physical region not only because of their 1\arcsec~resolution, but also because the emitting region of synchrotron self-absorbed emission shrinks with frequency as $R_\mathrm{pc} \sim \nu^{-7/4}_\mathrm{GHz}$ \citep{Laor2008,Behar2018}.

\subsection{The Fundamental Plane of Black Hole Activity}
\label{sec:fundplane_disc}

Section~\ref{sec:fundplane} discusses how our 22~GHz radio core luminosities, X-ray luminosities, and black hole masses compare to the black hole fundamental planes of \citet{Merloni2003} and by \citet{Bonchi2013}, who used a sample more similar to ours. Figure~\ref{fig:bonchi_merloni_planes} shows that the BAT AGN sample falls significantly below the \citet{Merloni2003} plane, and below the \citet{Bonchi2013} plane to a lesser degree. Although the core properties of objects with extended star formation in radio images fall the farthest below the fundamental plane predictions as a group, there is no apparent difference in adherence to the fundamental plane by extended radio morphology.

The following considerations should be borne in mind when viewing the comparisons between our sample and canonical fundamental planes (Figure~\ref{fig:bonchi_merloni_planes}). The mechanism responsible for radio emission at 22~GHz is potentially different than at 5~GHz; high frequency emission may represent a distinct spectral component \citep{Antonucci1988}. In star-forming galaxies, 22~GHz is the regime in which free-free emission becomes important in relation to pure synchrotron. There is also the possibility discussed above, that the 22~GHz emission is probing a nuclear coronal region beyond the reach of lower frequencies: since the synchrotron self-absorption coefficient scales as $\nu^{-3}$, higher frequencies can probe smaller emitting regions \citep[e.g.,][]{Laor2008}. A further important consideration may be spectral aging: if the magnetic field is unchanging, the rate of synchrotron losses is proportional to the square of the frequency, leading to higher frequencies dimming more quickly if further particle acceleration is not ongoing \citep{Jaffe1973, Harwood2013}. In a coronal situation, particle acceleration may be ongoing, but if the core emission contains a jet component then aging may affect the relative strengths of the 5~GHz and 22~GHz emission.

If aging is indeed affecting the relative strengths, the fact that the BAT AGN 22~GHz luminosities lie below the fundamental plane prediction may mean that most of the sources are ``turning off.'' This may indicate that samples compiled based on radio detection (e.g., Merloni's, but by definition not ours) have relatively short radio lifetimes. 

When \citet{Merloni2003} first unified black hole mass with radio and X-ray luminosity in the fundamental plane, the claim was that such a correlation across a huge range of masses implied an accretion flow/jet geometry that is scale invariant, depending only upon the accretion rate relative to the Eddington rate. In order to ensure the universality of the relation, this work included a large diversity of AGN subclasses, including very low accretion rate objects that we might consider quiescent, Seyferts of both optical types, and quasars. 

At the same time, \citet{Falcke2004} investigated whether radio-loud quasars, radio galaxies, and blazars are the supermassive analogs of the X-ray binary ``low-hard" state, in which the disk has receded and the jet is powerful. They found that after accounting for black hole mass, the scaling between radio and X-ray core luminosities followed a correlation reaching from X-ray binaries to FR~I radio galaxies, low-luminosity AGN, and BL Lac objects and followed jet-based scaling relations. Much more recently, \citet{Saikia2018} performed a similar study in a sample of low-luminosity AGN with sub-arcsecond radio resolution at 15~GHz, and obtained a slightly different fundamental plane from \citet{Merloni2003} that extends to AGN from X-ray binaries in the low-hard state. 

There is significant debate over what objects should be included in the fundamental plane, but the general opinion is that only supermassive black holes analogous to the low-hard state of X-ray binaries, and therefore with X-ray and radio emission generated by jet physics, should follow the relation. Objects with an expected significant coronal contribution (i.e., Seyferts with a standard accretion disk) would increase the intrinsic scatter in the relationship, as has been found by \citet{Kording2006} and \citet{Gultekin2009b}. The statistical differences between the many samples that have been used to construct fundamental planes and their physical implications are very well reviewed by \citet{Plotkin2012}. Despite the discrepancies of our sample from past fundamental planes when focusing only on AGN, they do not deviate on the grand scale of accretion spanning from stellar mass black holes (Figure~\ref{fig:plotkin}). They do indeed exhibit greater scatter, as expected for Seyferts with significant contributions from the corona and in contrast to the low-luminosity AGN in the figure, in which emission is likely to be dominated by jets.  It is also known that the scatter around the fundamental plane is enhanced when a broad sample of accretion rates are considered, as opposed to low-accretion-rate objects only \citep{Kording2006,Plotkin2012}.

\citet{Wong2016} conducted a preliminary analysis of the 1.4~GHz radio emission of the BASS AGN using FIRST and NVSS data. In contrast to the new 22~GHz observations, they did not find a significant offset of the 1.4~GHz to soft X-rays fundamental plane between the BASS sample relative to \citet{Merloni2003}.  However, it should be noted that our sample here is approximately four times larger than that of \citet{Wong2016}.  
In general, we expect greater sensitivity to jet age at 22~GHz \citep{Jaffe1973}; high radio frequencies and ultra-hard X-ray luminosities may also vary more rapidly since they are likely to originate in physically smaller regions. Both could result in the greater scatter observed in $L_\mathrm{22GHz}$ and in $L_\mathrm{14-195keV}$, relative to $L_\mathrm{1.4GHz}$ and $L_\mathrm{2-10keV}$, respectively. Since the estimation of $L_\mathrm{2-10keV}$ in \citet{Wong2016} was made by scaling the observed $L_\mathrm{14-195keV}$, the differences observed between these results can be primarily attributed to the new high angular resolution 22~GHz observations presented in this study. 

In addition to being larger and higher resolution than the \citet{Wong2016} data set, our sample differs from other previous ones in a few important ways: the \citet{Falcke2004} and \citet{Saikia2018} investigations target a particular model for jet-dominated states with inefficient accretion flows that may not form a disk, focusing on low-luminosity AGN or low-hard state analogs at low Eddington ratios. The \citet{Merloni2003} sample was very broad in type, consisting of data from different radio observatories and at different resolutions (albeit to a lesser degree than the \citealt{Falcke2004} sample). We note that these investigations were appropriate to their respective goals. Our sample, however, is appropriate for studying the core properties of AGN without pronounced radio jets, and was not selected to be low-accretion rate and is not likely to be dominated by objects analogous to the low-hard state of X-ray binaries. 

Figure~\ref{fig:bonchi_merloni_planes} also shows the fundamental plane relations with Eddington ratio color scaling. Interestingly, objects with low Eddington ratios, and therefore more likely low-hard state analogs, do not adhere more closely to the fundamental planes. In fact, for a given value of predicted radio flux (and therefore a given value of $\xi_{\mathrm{RX}}~\mathrm{log}~L_{X} + \xi_{\mathrm{RM}}~\mathrm{log}~M$, objects with higher Eddington ratios have higher observed 22~GHz luminosities closer to the fundamental plane predictions.

\subsection{Predicting M$_\mathrm{BH}$ with the Fundamental Plane}

Given the apparent universality of the fundamental plane, it is tempting to use the readily-observable and often archival quantities of $L_R$ and $L_X$ to estimate black hole mass. Such an effort is, unfortunately, plagued with complexity. The recent effort by \citet{Gultekin2019} is quite sophisticated, and is calibrated using only well-determined AGN black hole masses from dynamical estimates (as well as X-ray binaries). We refer the reader to their comprehensive discussion. The upshot is that the fundamental plane is an imprecise black hole mass estimator with large scatter. The exact coefficients depend upon which subsamples are included; building on the work of \citet{Gultekin2009b}, \citet{Gultekin2019} could not say definitively whether or not AGN and XRBs belong on the same fundamental plane. The AGN-only relation is quite different from the fit including XRBs; they ascribe this to possible physical differences in the relationship between X-ray and radio emission mechanisms. The subset of AGN referred to by \citet{Gultekin2019} as ``radio-active Seyferts" is most similar to our sample, and are consistent in that work with the fundamental plane derived for low luminosity AGN that are likely to be analogous to the XRB low-hard state. We note that \citet{Gultekin2019} state that the X-ray and radio flux measurements should be as close to simultaneous as possible for black hole mass prediction.

\begin{figure}
    \centering
    \includegraphics[width=\textwidth]{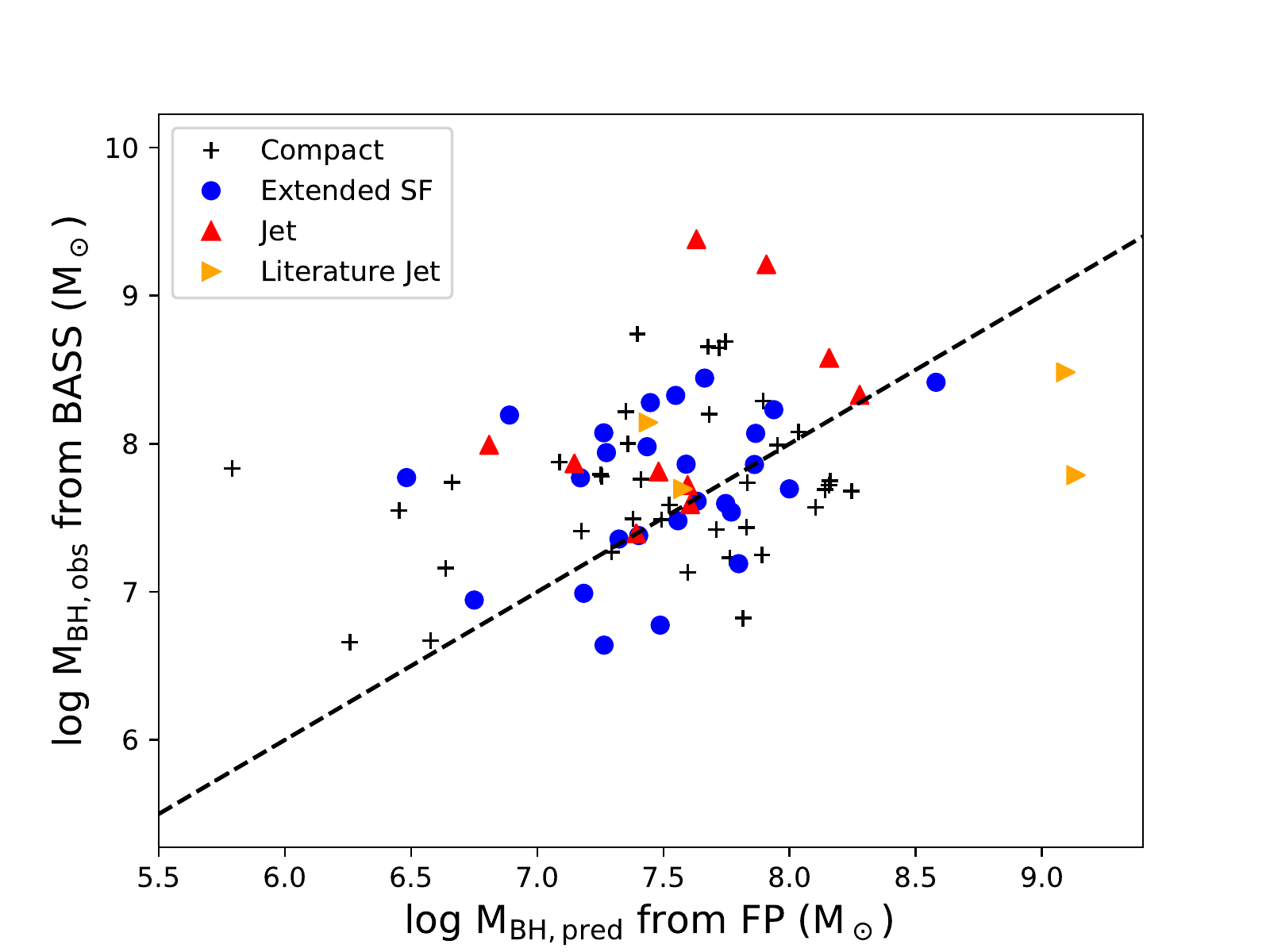}
    \caption{Comparison of the BAT sample's measured black hole masses and predicted black hole mass from fundamental plane parameters $L_X$~ (2-10~keV) and $L_R$~ (5~GHz) using the recommended relation from \citet{Gultekin2019}. The dashed line indicates the 1-to-1 relation.}  
        \label{fig:gult}
\end{figure}

In Figure~\ref{fig:gult} we show the results of using our 2-10~keV luminosities and estimated 5~GHz luminosities (Section~\ref{sec:fundplane}) in the recommended mass prediction equation from \citet{Gultekin2019}. The measurements are consistent with the predictions, but with large scatter, like those of \citet{Gultekin2019}. Our scatter is slightly larger, which is expected due to large time disparity of our data, our variety of black hole mass estimation methods, and the fact that our 5~GHz measurements are estimated and not measured at that frequency.

Our results agree with the conclusion of \citet{Gultekin2019} and other studies \citep[e.g., ][]{Nisbet2016}, that the fundamental plane is a poor predictor of black hole mass, only useful if other methods are impossible, or to distinguish broadly between XRBs, intermediate mass black hole candidates, and AGN.

\section{Conclusions}
\label{sec:conclusion}

We have presented the results of a 22~GHz 1\arcsec~resolution imaging survey of 100 radio-quiet, ultra-hard X-ray selected AGN, 70 of which were previously studied by \citet{Smith2016}, and 87 of which now have optical spectroscopic follow-up from the BASS collaboration. As in the previous phase of the survey, the observed morphologies fall broadly into three categories: compact/unresolved, extended and patchy emission indicative of star formation, and jet-like. After isolating the core emission (which encompasses spatial regions from $\sim60$~pc to $\sim1$~kpc for the sample redshifts), we compare it to predictions for the relationship between radio luminosity, X-ray luminosity, and black hole mass from the literature. Our conclusions are as follows:

1. Out of 100 radio-quiet AGN, mostly of the Seyfert class, 96 are radio-detected, and all 96 have a compact core in addition to any extended emission. This detection fraction of radio cores is very high compared to other surveys.

2. The ratio of total radio to X-ray luminosity, hard or ultra-hard, does not differentiate between kiloparsec-scale jets and nuclear star formation as the origin of radio emission. It is therefore very risky to use $L_R / L_X$~ as a diagnostic tool to distinguish between jets and star formation in low-resolution radio surveys.

3. Objects with kiloparsec-scale radio jets are more likely to have higher X-ray luminosities (both hard and ultra-hard) than objects with compact or star formation-dominated 22~GHz morphologies, potentially indicating a contribution to the X-ray luminosity from the jet in addition to the corona. 

4. The $L_R/L_X$ relationship is consistent with the \citet{Guedel1993} correlation for coronally-active stars; more so than previous AGN samples for which this has been investigated. We postulate that this is because the ultra-hard, absorption-corrected X-ray emission in our sample is more representative of nuclear power than previous samples, and perhaps because our 22~GHz flux measurements are especially well-suited to studying coronal emission. Further, the core properties of jetted objects are the best match to the coronal prediction.

5. The fundamental plane of black hole activity as put forward by \citet{Merloni2003} over-predicts our core radio emission, even when our 22~GHz measurements are corrected to 5~GHz using standard scaling or interpolation. The fundamental plane from \citet{Bonchi2013} is a much better match and was constructed from a sample much more similar to this one (hard X-ray selected, with 1\arcsec~radio imaging and excluding extended emission), but still over-predicts the core radio emission for objects with star formation morphologies in particular.

Finally, we note some ongoing developments in our own collaboration. \citet{Baek2019} has obtained sub-arcsecond resolution 22~GHz imaging with the Korean VLBI Network for 10 BAT AGN. Their sample is much more radio-loud than ours, with typical $L_R / L_X$ of $10^{-2}$, largely because it was selected to be bright enough for VLBI imaging in their fringe survey. Their preliminary work finds much better agreement with the existing fundamental planes than the results presented in this paper, which is consistent with the paradigm presented in Section~\ref{sec:fundplane_disc} in which objects that are analogous to the low-hard state (e.g., jet-dominated as opposed to corona- or disk-dominated Seyferts) are closer to the fundamental plane. The \citet{Baek2019} sample's far higher resolution may also contribute to less scatter about the fundamental plane, as it would naturally exclude any extended emission that remains unresolved within our 1\arcsec~beam.

The third and final phase of the 22~GHz survey consists of 128 objects to complete the sample observed by \emph{Herschel} and visible from the JVLA. We expect the full survey to be complete and published by approximately 2022. This largest and final iteration has no selection criteria regarding radio loudness and will therefore encompass a much larger spread in radio luminosity and include objects with classical radio jets.

\section*{Acknowledgements}
Support for KLS was provided by the National Aeronautics and
Space Administration through Einstein Postdoctoral Fellowship
Award Number PF7-180168, issued by the Chandra
X-ray Observatory Center, which is operated by the
Smithsonian Astrophysical Observatory for and on behalf
of the National Aeronautics Space Administration
under contract NAS8-03060.  CR acknowledges the CONICYT+PAI Convocatoria Nacional subvencion a instalacion en la academia convocatoria a\~{n}o 2017 PAI77170080. KO is an International Research Fellow of the Japan Society for the Promotion of Science (JSPS) (ID:P17321). FR acknowledges support from FONDECYT Postdoctorado 3180506; FR and FEB acknowledge support from CONICYT project Basal AFB-170002.

\small
\bibliographystyle{mnras}
\bibliography{biblio_bassvla_rev_mnras}

\renewcommand{\arraystretch}{2}
\onecolumn
\begin{center}
\scriptsize
\begin{longtable}{lccccccccccc}

\caption{Measured Parameters of the 22~GHz BAT AGN Sample}
\label{t:tab1}\\
\hline
\hline
Name & $z$ & Sy & $S_{\nu,\mathrm{1"}}$  &  $S_{\nu,\mathrm{6"}}$  & Morph. & log $L_\mathrm{HX,obs}$ & log $L_\mathrm{HX,int}$ & log $L_\mathrm{UHX,obs}$ & log $L_\mathrm{UHX,int}$ & Log $M_{BH}$ & $L/L_\mathrm{Edd}$  \\
  &  & Type & mJy  &  mJy  & 22~GHz & (erg s$^{-1}$)  & (erg s$^{-1}$) & (erg s$^{-1}$) & (erg s$^{-1}$) & $M_\odot$  &   \\
\hline

\endfirsthead
\multicolumn{11}{c}%
{\tablename\ \thetable\ -- \textit{Continued from previous page}} \\
\hline
\hline
Name & $z$ & Sy & $S_{\nu,\mathrm{1"}}$  &  $S_{\nu,\mathrm{6"}}$  & Morph. & log $L_\mathrm{HX,obs}$ & log $L_\mathrm{HX,int}$ & log $L_\mathrm{UHX,obs}$ & log $L_\mathrm{UHX,int}$ & Log $M_{BH}$ & $L/L_\mathrm{Edd}$  \\
  &  & Type & mJy  &  mJy  & 22~GHz & (erg s$^{-1}$)  & (erg s$^{-1}$) & (erg s$^{-1}$) & (erg s$^{-1}$) & $M_\odot$  &   \\
\hline

\endhead
\hline \multicolumn{11}{r}{\textit{Continued on next page}} \\
\endfoot
\hline
 \\
\caption{Properties of the 22~GHz \emph{Swift}-BAT AGN sample. Columns are (1) object name, (2) redshift, (3) 22~GHz flux density in the 1\arcsec~core, (4) 22~GHz flux density including extended emission, (5) morphology of the 22~GHz emission where C means compact, J means jet-like, E means extended but nonlinear, and LJ means the object has a known radio jet at lower frequencies in the literature but is compact in our sample, (6) $2-10$~keV luminosity as observed, (7) intrinsic $2-10$~keV luminosity corrected for absorption, (8) $14-195$~keV luminosity as observed, (9) intrinsic $14-195$~keV luminosity corrected for absorption, (10) black hole mass, and (11) Eddington ratio. Objects without values in some X-ray luminosity columns are not yet included in BASS (see Section~\ref{sec:bass}) and so do not have absorption correction; all objects have obesrved $14-195$~keV luminosities from the \emph{Swift}-BAT survey itself \citep{Baumgartner2013}. Entries with X-ray luminosity values but without black hole masses or Eddington ratio do not have confirmed black hole mass measurements in BASS. Objects with asterisks following their names were observed for approximately twice as long in our campaign to uncover very low surface-brightness star formation as described in Section~\ref{sec:radioprocessing}. }
\endlastfoot
2MASX J0025+6821&	0.012	&	2	&	1.18	&	1.17	&	C	&	41.35	&	43.14	&	42.77	&	43.06	&	$7.87^{+0.11}_{-0.12}$	&	0.005	\\
2MASX J0353+3714*	&	0.019	&	2	&	0.28	&	0.80	&	E	&	42.45	&	42.53	&	43.06	&	43.06	&	$7.19^{+0.1}_{-0.1}$	&	0.047	\\
2MASX J0423+0408	&	0.046	&	2	&	0.58	&	6.99	&	J	&	42.9	&	43.82	&	44.04	&	44.22	&		&		\\
2MASX J0444+2813*	&	0.01	&	2	&	3.05	&	3.20	&	C	&	42.55	&	42.67	&	43.18	&	43.12	&	$7.79^{+0.07}_{-0.07}$	&	0.015	\\
2MASX J0505-2351	&	0.036	&	2	&	1.72	&	2.18	&	C	&	43.33	&	43.48	&	44.23	&	44.21	&	$7.77^{+0.11}_{-0.11}$	&	0.181	\\
2MASX J1200+0648	&	0.036	&	2	&	0.84	&	1.10	&	E	&	43.18	&	43.38	&	43.79	&	43.79	&	$8.27^{+0.05}_{-0.05}$	&	0.021	\\
2MASX J1546+6929	&	0.038	&	2	&	0.27	&	1.48	&	J	&	42.52	&	43.08	&	43.66	&	43.72	&	$8.57^{+0.04}_{-0.04}$	&	0.008	\\
2MASX J1937-0613	&	0.01	&	1.5	&	5.16	&	8.40	&	E	&	42.74	&	42.75	&	42.74	&	42.76	&	$6.64^{+0.04}_{-0.05}$	&	0.080	\\
2MASX J2010+4800	&	0.025	&	2	&	0.16	&	0.25	&	C	&	42.2	&	42.42	&	43.28	&	43.28	&	$7.72^{+0.05}_{-0.06}$	&	0.023	\\
2MFGC 02280	&	0.015	&	2	&	0.40	&	1.67	&	E	&	41.41	&	43.02	&	43.15	&	43.42	&	$7.35^{+0.13}_{-0.14}$	&	0.039	\\
ARK 347*	&	0.023	&	2	&	0.42	&	0.30	&	E	&	42.43	&	42.9	&	43.52	&	43.56	&	$8.19^{+0.03}_{-0.03}$	&	0.013	\\
ARP 102B	&	0.024	&	2	&	219.02	&	226.64	&	LJ	&	42.81	&	42.82	&	43.38	&	43.36	&	$7.78^{+0.1}_{-0.11}$	&	0.025	\\
ARP 151	&	0.021	&	1.2	&	0.52	&	0.55	&	C	&	43.01	&	43.02	&	43.29	&	43.28	&	$6.67^{+0.03}_{-0.02}$	&	0.265	\\
CGCG 122-055	&	0.022	&	1.5	&	1.79	&	2.03	&	C	&	42.41	&	42.43	&	43.15	&	43.08	&	$7.13^{+0.03}_{-0.03}$	&	0.066	\\
CGCG 229-015	&	0.027	&	1	&	0.20	&	0.41	&	C	&		&		&		&		&		&		\\
CGCG 300-062	&	0.033	&	2	&	0.24	&	0.47	&	C	&		&		&		&		&		&		\\
CGCG 312-012	&	0.026	&	2	&	0.70	&	0.99	&	C	&	42.13	&	42.38	&	43.13	&	43.08	&	$8.21^{+0.03}_{-0.03}$	&	0.005	\\
CGCG 420-015	*&	0.03	&	2	&	0.83	&	1.17	&	C	&	42.38	&	44	&	43.75	&	43.99	&	$8.74^{+0.07}_{-0.07}$	&	0.006	\\
CGCG 493-002	&	0.024	&	1.5	&	1.18	&	1.43	&	E	&		&		&		&		&		&		\\
ESO 511-G030	&	0.023	&	1	&	12.24	&	12.04	&	C	&	43.34	&	43.47	&	43.66	&	43.66	&	$7.23^{+0.05}_{-0.05}$	&	0.171	\\
ESO 548-G081	&	0.014	&	1	&	0.46	&	2.60	&	E	&	43.01	&	43.01	&	43.32	&	43.33	&	$7.94^{+0.02}_{-0.02}$	&	0.015	\\
ESO 549- G 049	&	0.026	&	1.9	&	0.77	&	2.71	&	E	&	42.92	&	43.01	&	43.6	&	43.58	&	$8.07^{+0.07}_{-0.07}$	&	0.022	\\
IC 0486	&	0.027	&	1.9	&	0.77	&	1.80	&	E	&	42.8	&	42.81	&	43.72	&	43.69	&	$8.07^{+0.04}_{-0.04}$	&	0.028	\\
IC 2461	&	0.008	&	2	&	0.46	&	0.57	&	C	&	41.63	&	41.78	&	42.39	&	42.39	&	$7.26^{+0.1}_{-0.1}$	&	0.008	\\
IC 2637	&	0.029	&	1.5	&	2.01	&	5.32	&	E	&	42.7	&	42.7	&	43.38	&	43.32	&	$8.41^{+0.28}_{-0.22}$	&	0.006	\\
IGR J23308+7120*	&	0.037	&	2	&	0.11	&	0.51	&	C	&	42.65	&	42.88	&	43.55	&	43.5	&	$7.68^{+0.14}_{-0.15}$	&	0.047	\\
IRAS 05589+2828*	&	0.033	&	1.2	&	2.46	&	2.79	&	C	&	43.67	&	43.68	&	44.21	&	44.19	&	$8.69^{+0.22}_{-0.17}$	&	0.021	\\
LEDA 214543	&	0.032	&	2	&	1.33	&	1.34	&	C	&	42.95	&	43.07	&	43.76	&	43.73	&	$8.07^{+0.1}_{-0.1}$	&	0.030	\\
MCG -01-13-025	&	0.016	&	1.5	&	10.83	&	10.67	&	C	&	42.84	&	42.84	&	43.25	&	43.23	&		&		\\
MCG -01-24-012	&	0.02	&	2	&	4.81	&	5.12	&	LJ	&	43.05	&	43.24	&	43.55	&	43.55	&	$7.69^{+0.06}_{-0.06}$	&	0.046	\\
MCG-01-30-041	&	0.018	&	1.8	&	0.37	&	1.48	&	E	&		&		&		&		&		&		\\
MCG -01-40-001	&	0.023	&	1.9	&	12.73	&	23.00	&	J	&	43.07	&	43.24	&	43.58	&	43.58	&	$9.20^{+0.25}_{-0.29}$	&	0.001	\\
MCG -02-08-014	&	0.017	&	2	&	0.74	&	1.19	&	J	&	42.55	&	42.83	&	43.22	&	43.24	&	$7.86^{+0.05}_{-0.05}$	&	0.014	\\
MCG -02-12-050	&	0.036	&	1.2	&	0.64	&	1.52	&	C	&	43.32	&	43.39	&	43.77	&	43.74	&	$8.23^{+0.14}_{-0.11}$	&	0.022	\\
MCG -05-23-016	&	0.008	&	1.9	&	3.47	&	3.62	&	C	&	43.15	&	43.2	&	43.51	&	43.5	&		&		\\
MCG +02-57-002	&	0.03	&	1.9	&	0.38	&	0.53	&	E	&	42.64	&	42.6	&	43.43	&	43.39	&	$7.38^{+0.04}_{-0.04}$	&	0.071	\\
MCG +04-22-042	&	0.033	&	1.2	&	1.03	&	1.73	&	J	&	43.45	&	43.45	&	43.98	&	43.97	&	$7.59^{+0.04}_{-0.06}$	&	0.156	\\
MCG +04-48-002	&	0.014	&	2	&	0.44	&	8.97	&	E	&	42.05	&	43.13	&	43.52	&	43.44	&	$7.76^{+0.06}_{-0.06}$	&	0.036	\\
MCG +06-16-028	&	0.016	&	1.9	&	2.24	&	3.41	&	E	&	41.22	&	43.07	&	42.97	&	43.38	&		&		\\
MCG +08-11-011	&	0.02	&	1.5	&	13.84	&	15.85	&	J	&	43.62	&	43.79	&	44.1	&	44.1	&	$7.81^{+0.03}_{-0.04}$	&	0.124	\\
MCG +11-11-032	&	0.036	&	2	&	0.13	&	0.14	&	C	&	42.98	&	43.44	&	43.72	&	43.76	&	$8.28^{+0.06}_{-0.06}$	&	0.017	\\
Mrk 10	&	0.029	&	1.5	&	0.32	&	0.56	&	C	&	43.12	&	43.12	&	43.46	&	43.46	&	$7.25^{+0.1}_{-0.07}$	&	0.103	\\
Mrk 1392	&	0.036	&	1.5	&	0.45	&	1.56	&	E	&	43.11	&	43.11	&	43.75	&	43.72	&	$7.86^{+0.01}_{-0.01}$	&	0.049	\\
Mrk 18	&	0.011	&	1.9	&	3.45	&	5.13	&	E	&	41.58	&	41.82	&	42.52	&	42.52	&	$7.69^{+0.05}_{-0.05}$	&	0.004	\\
Mrk 198*	&	0.024	&	2	&	0.96	&	1.45	&	E	&	42.81	&	42.98	&	43.47	&	43.48	&	$7.86^{+0.05}_{-0.05}$	&	0.026	\\
Mrk 279	&	0.03	&	1.5	&	3.03	&	3.08	&	C	&	43.41	&	43.41	&	43.92	&	43.91	&	$7.43^{+0.09}_{-0.13}$	&	0.194	\\
Mrk 359	&	0.017	&	1.5	&	0.53	&	0.80	&	C	&	42.69	&	42.7	&	42.96	&	42.94	&		&		\\
Mrk 477	&	0.038	&	1.9	&	5.45	&	6.11	&	C	&	42.69	&	43.26	&	43.68	&	43.56	&		&		\\
Mrk 50	&	0.024	&	1	&	0.37	&	0.31	&	C	&	43.1	&	43.1	&	43.45	&	43.45	&	$7.42^{+0.01}_{-0.01}$	&	0.068	\\
Mrk 590*	&	0.027	&	1.5	&	2.02	&	2.74	&	C	&	42.7	&	42.7	&	43.42	&	43.39	&	$7.56^{+0.06}_{-0.09}$	&	0.045	\\
Mrk 728	&	0.036	&	1.5	&	1.32	&	1.30	&	C	&	43.03	&	43.02	&	43.6	&	43.55	&	$7.76^{+0.01}_{-0.01}$	&	0.044	\\
Mrk 739E	&	0.03	&	1	&	0.31	&	1.36	&	E	&	43.16	&	43.18	&	43.43	&	43.43	&	$6.99^{+0.02}_{-0.02}$	&	0.175	\\
Mrk 766	&	0.013	&	1.5	&	4.60	&	4.84	&	C	&	42.69	&	42.71	&	42.91	&	42.9	&	$6.82^{+0.08}_{-0.08}$	&	0.078	\\
Mrk 79	&	0.022	&	1.5	&	1.45	&	2.39	&	E	&	42.93	&	43.11	&	43.72	&	43.7	&	$7.61^{+0.02}_{-0.03}$	&	0.081	\\
Mrk 817	&	0.031	&	1.2	&	1.94	&	2.10	&	C	&	43.49	&	43.49	&	43.77	&	43.77	&	$7.58^{+0.02}_{-0.03}$	&	0.097	\\
Mrk 885	&	0.025	&	1.5	&	0.23	&	0.30	&	C	&		&		&		&		&		&		\\
Mrk 926	&	0.047	&	1.5	&	8.62	&	9.77	&	C	&	44.18	&	44.18	&	44.77	&	44.75	&	$7.99^{+0.05}_{-0.05}$	&	0.383	\\
Mrk 975	&	0.049	&	1.2	&	1.26	&	1.55	&	C	&	43.3	&	43.56	&	43.98	&	43.97	&	$7.75^{+0.04}_{-0.05}$	&	0.108	\\
NGC 1052	&	0.004	&	1.9	&	1010.50	&	1083.00	&	LJ	&	41.46	&	41.62	&	42.21	&	42.18	&	$8.48^{+0.03}_{-0.03}$	&	3.3e-4	\\
NGC 1106	&	0.014	&	2	&	11.15	&	11.62	&	C	&		&		&		&		&		&		\\
NGC 1125	&	0.011	&	2	&	6.14	&	6.42	&	C	&	41.03	&	42.74	&	42.67	&	42.96	&	$7.49^{+0.07}_{-0.07}$	&	0.010	\\
NGC 1194	&	0.014	&	2	&	1.08	&	1.26	&	C	&	41.62	&	43.69	&	43.18	&	43.68	&	$7.83^{+0.04}_{-0.04}$	&	0.014	\\
NGC 2110	&	0.007	&	2	&	42.17	&	66.10	&	J	&	42.52	&	42.69	&	43.63	&	43.63	&	$9.37^{+0.07}_{-0.07}$	&	0.001	\\
NGC 235A	&	0.022	&	1.9	&	3.28	&	4.33	&	E	&	42.65	&	43.21	&	43.72	&	43.76	&	$8.44^{+0.03}_{-0.03}$	&	0.012	\\
NGC 2655	&	0.005	&	2	&	12.51	&	12.28	&	C	&	40.73	&	41.36	&	41.82	&	41.87	&	$8.19^{+0.1}_{-0.11}$	&	2.6e-4	\\
NGC 2992	&	0.008	&	1.9	&	12.49	&	29.27	&	E	&	41.98	&	42	&	42.55	&	42.52	&	$8.32^{+0.14}_{-0.15}$	&	0.001	\\
NGC 3081	&	0.008	&	2	&	1.21	&	1.62	&	C	&	41.55	&	42.72	&	43.07	&	43.29	&	$7.73^{+0.06}_{-0.06}$	&	0.014	\\
NGC 3227	&	0.003	&	1.5	&	4.13	&	6.95	&	E	&	42.09	&	42.1	&	42.57	&	42.55	&	$6.77^{+0.03}_{-0.03}$	&	0.040	\\
NGC 3431	&	0.017	&	2	&	0.69	&	0.97	&	C	&	42.22	&	42.39	&	43.19	&	43.13	&	$7.73^{+0.06}_{-0.06}$	&	0.018	\\
NGC 3516	&	0.009	&	1.2	&	3.70	&	5.28	&	J	&	42.67	&	42.72	&	43.31	&	43.29	&	$7.39^{+0.06}_{-0.04}$	&	0.052	\\
NGC 3718	&	0.003	&	1.9	&	17.99	&	18.63	&	LJ	&	40.58	&	40.61	&	41.46	&	41.47	&	$8.14^{+0.11}_{-0.12}$	&	1.3e-4	\\
NGC 3786	&	0.009	&	1.9	&	0.72	&	2.06	&	E	&	42.06	&	42.11	&	42.5	&	42.45	&	$7.47^{+0.16}_{-0.18}$	&	0.007	\\
NGC 4235	&	0.008	&	1.2	&	12.33	&	12.60	&	C	&	41.6	&	41.6	&	42.66	&	42.64	&		&		\\
NGC 4388	&	0.008	&	2	&	3.26	&	12.71	&	E	&	42.5	&	43.05	&	43.64	&	43.7	&	$6.94^{+0.13}_{-0.14}$	&	0.315	\\
NGC 513*	&	0.019	&	2	&	0.87	&	0.68	&	E	&	42.52	&	42.66	&	43.24	&	43.22	&	$7.53^{+0.12}_{-0.12}$	&	0.032	\\
NGC 5231	&	0.022	&	2	&	0.64	&	1.11	&	C	&	42.81	&	42.89	&	43.22	&	43.16	&	$8.00^{+0.04}_{-0.04}$	&	0.011	\\
NGC 5273	&	0.004	&	1.5	&	0.55	&	0.68	&	C	&	41.22	&	41.26	&	41.57	&	41.48	&	$6.65^{+0.13}_{-0.19}$	&	0.005	\\
NGC 5290	&	0.009	&	2	&	6.99	&	8.08	&	C	&	41.91	&	41.93	&	42.5	&	42.46	&	$7.78^{+0.06}_{-0.06}$	&	0.003	\\
NGC 5506	&	0.006	&	1.9	&	48.53	&	48.61	&	C	&	42.9	&	42.99	&	43.31	&	43.3	&		&		\\
NGC 5548	&	0.017	&	1.5	&	1.44	&	4.52	&	J	&	43.1	&	43.14	&	43.72	&	43.7	&	$7.71^{+0.11}_{-0.05}$	&	0.064	\\
NGC 5683	&	0.037	&	1.2	&	0.39	&	0.49	&	C	&	43.1	&	43.07	&	43.57	&	43.55	&	$7.69^{+0.01}_{-0.02}$	&	0.048	\\
NGC 5728	&	0.01	&	1.9	&	4.08	&	7.88	&	J	&	41.43	&	42.86	&	43.23	&	43.36	&	$7.99^{+0.07}_{-0.07}$	&	0.011	\\
NGC 6552	&	0.026	&	2	&	4.76	&	5.35	&	C	&		&		&		&		&		&		\\
NGC 7679	&	0.017	&	2	&	0.46	&	7.22	&	E	&		&		&		&		&		&		\\
NGC 788	&	0.014	&	2	&	0.61	&	1.07	&	E	&	42.12	&	43.02	&	43.52	&	43.66	&	$7.77^{+0.11}_{-0.12}$	&	0.036	\\
NGC 931	&	0.016	&	1.5	&	0.93	&	1.24	&	C	&	43.25	&	43.41	&	43.58	&	43.58	&	$7.41^{+0.06}_{-0.07}$	&	0.094	\\
NGC 985	&	0.043	&	1.5	&	1.01	&	1.41	&	E	&	43.78	&	43.78	&	44.14	&	44.12	&	$7.98^{+0.02}_{-0.02}$	&	0.092	\\
SBS 1301+540	&	0.03	&	1.5	&	0.88	&	1.05	&	C	&	43.72	&	43.73	&	43.82	&	43.8	&	$7.55^{+0.02}_{-0.02}$	&	0.118	\\
UGC 03478	&	0.012	&	1.2	&	0.97	&	1.38	&	C	&		&		&		&		&		&		\\
UGC 03601*	&	0.017	&	1.9	&	1.26	&	1.58	&	E	&	42.66	&	42.67	&	43.14	&	43.14	&		&		\\
UGC 07064	&	0.025	&	1.9	&	0.61	&	1.15	&	E	&	42.53	&	42.58	&	43.27	&	43.15	&	$7.59^{+0.05}_{-0.05}$	&	0.030	\\
UGC 08327 NED02	&	0.035	&	2	&	2.57	&	2.96	&	C	&	43.34	&	43.57	&	43.72	&	43.73	&	$8.64^{+0.04}_{-0.04}$	&	0.008	\\
UGC 11185 NED02	&	0.041	&	2	&	6.82	&	8.17	&	J	&	43.28	&	43.35	&	43.88	&	43.83	&	$8.32^{+0.1}_{-0.11}$	&	0.023	\\
UGC 12282	&	0.017	&	2	&	0.44	&	0.55	&	C	&	41.76	&	42.62	&	43.09	&	43.23	&	$8.65^{+0.04}_{-0.04}$	&	0.002	\\
UGC 12741*	&	0.017	&	2	&	0.31	&	0.70	&	C	&	41.87	&	42.94	&	43.13	&	43.31	&	$7.48^{+0.12}_{-0.12}$	&	0.028	\\
UM 614	&	0.033	&	1.5	&	0.23	&	0.27	&	C	&	43.18	&	43.19	&	43.6	&	43.61	&	$7.16^{+0.01}_{-0.02}$	&	0.175	\\
\end{longtable}
\end{center}


\end{document}